\documentclass[prd,aps,showpacs,floats,floatfix,superscriptaddress]{revtex4}

\usepackage[dvips]{graphicx}

\usepackage{longtable}
\usepackage{dcolumn,epsfig}

\def\laq{\raise 0.4ex\hbox{$<$}\kern -0.8em\lower 0.62ex\hbox{$\sim$}}
\def\gaq{\raise 0.4ex\hbox{$>$}\kern -0.7em\lower 0.62ex\hbox{$\sim$}}

\newcommand{\beq}{\begin{equation}}
\newcommand{\eeq}{\end{equation}}
\newcommand{\bea}{\begin{eqnarray}} 
\newcommand{\eea}{\end{eqnarray}}
\newcommand{\ba}{\begin{array}}
\newcommand{\ea}{\end{array}}

\newcommand{\eqref}[1]{(\ref{#1})}

\newlength{\sizeonefig}
\newlength{\sizetwofig}
\newlength{\sizeonefigb}
\newlength{\sizetwofigb}
\begin{document}

\title{Scaling law in signal recycled laser-interferometer 
gravitational-wave detectors}

\author{Alessandra Buonanno} 

\affiliation{Institut d'Astrophysique de Paris (GReCO, FRE 2435 du CNRS), 
98$^{\rm bis}$ Boulevard Arago, 
75014 Paris, France}

\affiliation{Theoretical Astrophysics, California 
Institute of Technology, Pasadena, CA 91125}

\author{Yanbei Chen} 

\affiliation{Theoretical Astrophysics, California 
Institute of Technology, Pasadena, CA 91125}

\begin{abstract}
By mapping the signal-recycling (SR) optical configuration to a 
three-mirror cavity, and then to a single detuned cavity, we express 
SR optomechanical dynamics, input--output relation and noise spectral 
density in terms of \emph{only} three characteristic parameters: 
the (free) optical resonant frequency and decay
time of the entire interferometer, and the laser power circulating in arm cavities. 
These parameters, and therefore the properties of the interferometer,
are invariant under an appropriate scaling of SR-mirror reflectivity,
SR detuning, arm-cavity storage time and input power at beamsplitter. 
Moreover, so far the quantum-mechanical
description of laser-interferometer gravitational-wave detectors,  
including radiation-pressure effects, was only obtained at linear order 
in the transmissivity of arm-cavity internal mirrors. We relax this assumption and 
discuss how the noise spectral densities change.
\end{abstract}

\maketitle

\section{Introduction}
\label{sec1}

\begin{figure}
\begin{center} 
\epsfig{file=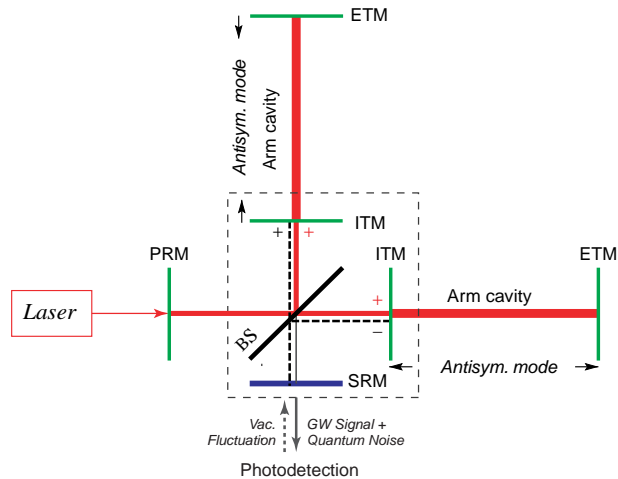,width=\sizeonefig}
\caption{\label{LIGOII} We draw a signal- (and power-) recycled LIGO
interferometer. The laser light enters the interferometer from the
left (bright port), through the power-recycling mirror (PRM), and
get split by a 50/50 beamsplitter (BS) into the two identical (in 
absence of gravitational waves) arm cavities. Each of the arm cavities 
is formed by the internal test-mass mirror
(ITM) and the end test-mass mirror (ETM). No light leaves the
interferometer from below the BS (dark port), except the lights induced
by the antisymmetric motion of the test-mass mirrors, e.g., due to a
passing-by gravitational wave,  or due to vacuum fluctuations that 
originally enter the interferometer from the dark port. A
SRM is placed at the dark port, forming a
SR cavity (marked by thick dashed lines) with the ITMs.}
\end{center}
\end{figure}

A network of broadband ground-based laser interferometers,
aimed at detecting gravitational waves (GWs) in the frequency band
$10-10^4\,$Hz, is already operating. 
This network is composed of GEO, the Laser Interferometer 
Gravitational-wave Observatory (LIGO), TAMA and VIRGO 
(whose operation will begin in 2004)~\cite{Inter}.
The LIGO Scientific Collaboration (LSC) \cite{GSSW99} is 
currently planning an upgrade of LIGO starting from 2008, called  
advanced LIGO or LIGO-II. 
Besides the improvement of the seismic isolation and 
suspension systems, and the increase (decrease) of light power 
(shot noise) circulating in arm cavities, the LIGO community  
has planned to introduce an extra mirror, called a 
signal-recycling mirror (SRM)~\cite{D82,SRt}, 
at the dark-port output (see Fig.~\ref{LIGOII}). 
The optical system composed of SR cavity and arm 
cavities forms a composite resonant cavity, whose 
eigenfrequencies and quality factors can be controlled by
the position and reflectivity of the SR mirror.
These eigenfrequencies (resonances) can be exploited 
to reshape the noise curves, enabling the interferometer to work
either in broadband or in narrowband
configurations, and improving in this way
the observation of specific GW astrophysical
sources~\cite{KT}. 

The initial theoretical analyses~\cite{D82,SRt} and 
experiments~\cite{SRe} of SR interferometers 
refer to configurations with low laser power, 
for which the radiation pressure on the arm-cavity mirrors 
is negligible and the quantum-noise spectra are dominated by shot noise.
When the laser power is increased, the shot noise decreases 
while the effect of radiation-pressure fluctuation increases. 
LIGO-II has been planned to work at a laser power for which the
two effects are comparable in the observational band $40$--$200$\,Hz~\cite{GSSW99}. 
Thus, to correctly describe the quantum optical noise in LIGO-II, 
the results have been complemented by a thorough investigation of the influence of 
radiation-pressure force on mirror motion~\cite{KLMTV00,BC1,BC2,BC3}.
The analyses revealed that SR interferometers behave as 
an ``optical spring''. The dynamics of the whole optomechanical system, composed of 
arm-cavity mirrors and optical field, resembles 
that of a free test mass (mirror motion) connected to a massive spring (optical fields). 
When the test mass and the spring are not connected 
(e.g., for very low laser power) they have their own eigenmodes: the uniform
translation mode for the free mode and the 
longitudinal-wave mode for the spring. However, for LIGO-II laser power 
the test mass is connected to the massive spring and 
the two free modes get shifted in frequency, so the entire coupled
system can resonate at two pairs of finite frequencies.
Near these resonances the noise curve can 
beat the free mass standard quantum limit (SQL) for GW detectors~\cite{BK92}. Indeed, the SQL 
is not by itself an absolute limit, it depends on the dynamical properties 
of the test object (or probe) which we monitor.  
This phenomenon is not unique to SR interferometers; it 
is a generic feature of detuned cavities~\cite{All,FK} and 
was used by Braginsky, Khalili and colleagues in 
conceiving the ``optical bar'' GW detectors~\cite{OB}.
However, because the optomechanical system is by itself dynamically unstable, 
and a careful and precise study of the control system 
should be carried out~\cite{BC3}.  

The quantum mechanical analysis of SR interferometers given in Refs.~\cite{BC1,BC2,BC3},  
was built on results obtained by Kimble, Levin, Matsko, Thorne and Vyatchanin (KLMTV)~\cite{KLMTV00} 
for conventional interferometers, i.e.\ without SRM. For this reason, both the 
SR input--output relation~\cite{BC1,BC2} and the SR optomechanical dynamics~\cite{BC3} were 
expressed in terms of parameters characterizing conventional interferometers, such as 
the storage time in the arm cavities, instead of parameters
characterizing SR interferometers as a whole, 
such as the resonant frequencies and the storage time of the entire interferometer. 
Therefore, the analysis given in Refs.~\cite{BC1,BC2,BC3} is not fully suitable for highlighting 
the physics in SR interferometers. 

In this paper, we first map the SR interferometer into a three-mirror
cavity, as originally done by Mizuno~\cite{JM}, though in the low power limit and  
neglecting radiation-pressure effects, and by Rachmanov~\cite{MR} in classical regimes. 
Then, as first suggested by Mizuno~\cite{JM}, we regard the very short SR cavity 
(formed by SRM and ITM) as one (effective) mirror and we express input--output relation 
and noise spectral density~\cite{BC1}, and optomechanical 
dynamics~\cite{BC2} as well, in terms of three \emph{characteristic}
parameters that have more direct physical meaning: 
the {\it free} optical resonant frequency and decay time of the entire SR interferometer,   
and the laser power circulating in arm cavities. By free optical resonant 
frequency and decay time we mean the real and inverse imaginary part of 
the  (complex) optical resonant frequency when all the test-mass mirrors
are \emph{held fixed.} These parameters can then be represented in
terms of the more \emph{practical} parameters: the power transmissivity of ITM, 
the amplitude reflectivity of SRM, SR detuning and the input power. 
An appropriate scaling of the practical parameters can leave the 
characteristic parameters invariant. 

In addition, in investigating SR interferometers~\cite{BC1,BC2,BC3} 
the authors restricted the analyses to linear order in the transmissivity 
of arm-cavity internal mirrors, as also done by KLMTV~\cite{KLMTV00} for 
conventional interferometers. In this paper we relax this assumption and 
discuss how results change.

The outline of this paper is as follows. In Sec.~\ref{sec2} we explicitly  
work out the mapping between a SR interferometer and a three-mirror cavity, 
expressing the free oscillation frequency, decay time 
and laser power circulating in arm cavity, i.e.\ the characteristic parameters, 
in terms of SR-mirror reflectivity, SR detuning and arm-cavity storage time, 
which are the parameters used in the original
description~\cite{BC1,BC2}. An interesting scaling law among the practical
parameters is then obtained. In Secs.~\ref{sec3} and \ref{4.1} the input--output relations, noise spectral density 
and optomechanical dynamics are expressed in terms of those characteristic parameters.
In Sec.~\ref{4.2} we map the SR interferometer to a single detuned cavity 
of the kind analyzed by Khalili~\cite{FK}. In Sec.~\ref{4.3} 
we show that  correlations between shot noise and 
radiation-pressure noise in SR interferometers are equivalent to a change of the optomechanical 
dynamics, as discussed in a more general context by Syrtsev and Khalili~\cite{SK94}. 
In Sec.~\ref{4.4}, using fluctuation-dissipation 
theorem, we explain why optical spring detectors have \emph{very low} intrinsic noise, 
and are then preferable to mechanical springs in measuring very tiny forces.
In Sec.~\ref{sec5} we derive the input--output relation of SR interferometers 
at all orders in the transmissivity of internal test-mass mirrors. 
Finally, Sec.~\ref{sec6} summarizes our main conclusions. 
Appendix~\ref{quadraturerelations} contains definitions and 
notations, Appendix~\ref{stokes} discusses the Stokes relations 
in our optical system and in Appendix~\ref{secondorder}  we give 
the input--output relation including also next-to-leading order 
terms in the transmissivity of arm-cavity internal mirrors.

In this manuscript we shall be concerned only with quantum noise, though in realistic 
interferometers seismic and thermal noises are also present. Moreover, we shall 
neglect optical losses [see Ref.~\cite{BC2} where optical losses in 
SR interferometers were discussed].

\section{Derivation of scaling law}
\label{sec2}

\subsection{Equivalent three-mirror--cavity description of  signal-recycled interferometer}
\label{subsec2.1}

In Fig.~\ref{LIGOII}, we draw a signal- and power-
recycled LIGO interferometer. The Michelson-type optical
configuration makes it natural to decompose the optical fields and the
mechanical motion of the mirrors into modes that are either symmetric
(i.e.\ equal amplitude) or antisymmetric (i.e.\ equal in magnitude but opposite in signs) 
in the two arms, as done in Refs.~\cite{KLMTV00,BC1,BC2,BC3}, and  briefly explained 
in the following. In order to understand this decomposition more easily, 
let us for the moment ignore the power-recycling mirror (PRM) and
the signal-recycling mirror (SRM).  

First, let us suppose all mirrors are held fixed in their equilibrium 
positions. The laser light, which enters the interferometer 
from the left of the beamsplitter (BS), excites stationary, monochromatic carrier  
light inside the two identical arm cavities with equal amplitudes (marked with two +
signs in Fig.~\ref{LIGOII}) and thereby drives the symmetric
mode. To maximize the carrier amplitude inside the arm cavities, 
the arm lengths are chosen to be on resonance with the laser
frequency. When the carrier lights leave the two arms and recombine at 
the BS, they have the same magnitude and sign, and, 
as a consequence, leak out the interferometer only from the left port of the BS.  
No carrier light leaks out from the port below the BS. For this 
reason, the left port is called the bright port, and the
port below the BS is called the dark port.  Obviously, were there any other light that
enters the bright port, it would only drive the symmetric mode,
which would then leak out only from the bright port. Similarly, lights
that enter from the dark port would only drive the antisymmetric optical
mode, which have opposite signs at the BS (marked in
Fig.~\ref{LIGOII}) and would leak out the interferometer only from the dark port. 

Now suppose the mirrors (ITMs and ETMs) move in an antisymmetric
(mechanical) mode (shown by arrows in Fig.~\ref{LIGOII})  such that
the two arm lengths change in opposite directions --- for example driven by a
gravitational wave. This kind of motion would pump the
(symmetric) carriers in the two arms into sideband lights with opposite signs,
which lie in the antisymmetric mode, and would leak out the interferometer from
the dark port (and thus can be detected). On the contrary, symmetric mirror motions
that change the two arm lengths in the same way would induce sidebands
in the symmetric mode, which would leave the interferometer from the bright
port.  Moreover,  sideband lights inside the arm cavities, combined with the
strong carrier lights, exert forces on the test masses. Since the
carrier lights in the two arms are symmetric,  sidebands in the
symmetric (antisymmetric) optical mode drive only the symmetric
(antisymmetric) mechanical modes.  
In this way, we have two effectively decoupled systems in our
interferometer: (i) ingoing and outgoing bright-port optical fields, 
symmetric optical and mechanical modes, and (ii) 
ingoing and outgoing dark-port optical fields, antisymmetric optical
and mechanical modes.

When the PRM and SRM are present, since each of them only affects one
of the bright/dark ports, the decoupling between the symmetric and 
antisymmetric modes is still valid. 
Nevertheless, the behavior of each of the subsystems becomes richer. 
The PRM, along with the two ITMs, forms a power recycling
cavity (for symmetric optical modes, shown  by solid lines in 
Fig.~\ref{LIGOII}). In practice, in order to increase the carrier 
amplitude inside the arm cavities \cite{D82},  
this cavity is always set to be on resonance with the input laser light. More 
specifically, if the input laser power at the PRM is $I_{\rm in}$, then the power
input at the BS is $I_0=4 I_{\rm in}/T_{\rm p}$, and the circulating power
inside the arms is $I_c=2I_0/T$, where $T_{\rm p}$ and $T$ are the
power transmissivities of the PRM and the ITM. The SRM, along with
the two ITMs, forms a SR cavity (for the antisymmetric
optical modes, shown by dashed lines in Fig.~\ref{LIGOII}). 
By adjusting the length and finesse of this cavity, we can modify the resonant 
frequency and storage time of the antisymmetric
optical mode \cite{SRt}, and affect the optomechanical dynamics of
the entire  interferometer \cite{BC3}. These changes will reshape the
noise curves of SR interferometers, and can allow them
to beat the SQL~\cite{BC1,BC2}.

\begin{figure}
\begin{center} 
\epsfig{file=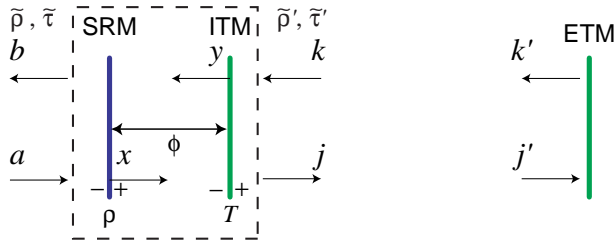,width=\sizeonefig}
\caption{\label{ITSRM} We draw the three-mirror cavity which is equivalent to a 
SR interferometer in describing the antisymmetric 
optical/mechanical modes and dark-port optical fields. The
SR cavity, which is mapped into a two-mirror cavity (in dashed
box) can be viewed as an effective mirror, with four effective
reflectivities and transmissivities, $\widetilde{\rho}'$,
$\widetilde{\tau}'$ (for fields entering from the right side),  and
$\widetilde{\rho}$, $\widetilde{\tau}$ (for fields entering from
the left side). The input and output fields, $a$ and $b$,
corresponds to those at the dark-port of the real  
SR interferometer. }
\end{center}
\end{figure}

Henceforth, we focus on the subsystem made up of dark-port fields and antisymmetric
optical and mechanical modes, in which the detected GW 
signal and quantum noises reside. In light of the above discussions,
it is convenient to identify the two arm cavities as one effective arm 
cavity, and map the entire interferometer to a three-mirror cavity, as shown in
Fig.~\ref{ITSRM}. In particular, the SR cavity, formed
by the SRM and ITMs  is mapped into a two-mirror  cavity (inside the
dashed box of Fig.~\ref{ITSRM}) or {\it one effective} ITM. 
The antisymmetric mechanical motions
of the two \emph{real} arm cavities is equal or opposite in sign to
those of this system. The input and output fields at the dark port corresponds
to those of the three-mirror cavity, $a$ and $b$ (shown in
Fig.~\ref{ITSRM}). Because of the presence of the BS in real
interferometer (and the absence in effective one), 
the optical fields inside the two real arms is 
$\pm 1/\sqrt{2}$ times the fields in the effective cavity composed of 
the effective ITM and ETM. As a consequence, fields in this effective cavity
are $\sqrt{2}$ times as sensitive to mirror motions as those in the real
arms, and the effective power in the effective cavity must be
\beq
\label{Iarm}
I_{\rm arm}=2I_c\,.
\eeq
Therefore, both the carrier amplitude and the
sideband amplitude in the effective cavity are $\sqrt{2}$ times
stronger than the ones in each real arm. In order to have the same effects on the
motion of the mirrors, we must impose the effective ETM and ITM to
be twice as massive as the real ones, i.e.
\beq
\label{marm}
m_{\rm arm}=2m\,.
\eeq
We denote by $T$ and $R=1-T$ the power transmissivity and reflectivity of 
the ITMs, $L = 4$ km is the arm length, and we assume the 
ETMs to be perfectly reflecting. The arm
length is on resonance with the carrier frequency $\omega_0 = 1.8 \times 10^{15}
\,{\rm sec}^{-1}$, i.e.\ $\omega_0 L/c=N\pi$, with $N$ an integer.  We denote 
by $\rho$ and $l$ the reflectivity of the SRM and the length of the 
SR cavity, and $\phi=[\omega_0l/c]_{\mathrm{mod}\,2\pi}$ 
the phase gained by lights with carrier
frequency  upon one trip across the SR cavity. We 
assume the SR cavity to be very short ($\sim 10$ m) 
compared with the arm-cavity length. Thus, we disregard the phase 
gained by lights with sideband frequency while traveling across 
the SR cavity, i.e.\ $\Omega\,l/c \rightarrow 0$.
The three-mirror cavity system can be broken 
into two parts. The effective arm cavity, which is the region to the right of 
the SR cavity, including the ETM (but {\it excluding} the ITM),
where the light interacts with the mechanical motion of the ETM. This
region is completely  characterized by the circulating
power $I_c$, the arm length $L$ and the mirror mass $m$. 
The (very short) SR cavity, made up of the SRM and the ITM,
which does not move. This part is characterized by $T$, $\rho$ and
$\phi$. 

Henceforth, we assume the radiation pressure
forces acting on the ETM and ITM to be equal, and the
contribution of the radiation-pressure--induced motion of the two
mirrors to the output light, or the radiation-pressure noises due to
the two mirrors, to be equal. [These assumptions introduce errors on the order
of $\max\{\Omega L/c,\,T\}$.] As a consequence, we can equivalently hold 
the ITM fixed and assume the ETM has a reduced mass of 
\beq
\label{muarm}
\mu_{\rm arm}=\frac{1}{2}m_{\rm arm}\,.
\eeq

\subsection{The scaling law in generic form}
\label{subsec2.2}

As first noticed by Mizuno~\cite{JM}, when the SR cavity is very short, 
we can describe it as a single effective mirror 
with  frequency-independent (but complex) effective transmissivities
and reflectivities (see Fig.~\ref{ITSRM}) $\widetilde{\rho}$, $\widetilde{\tau}$  
(for fields entering from the left) and $\widetilde{\rho}'$, $\widetilde{\tau}'$ 
(for fields entering from the right), and write the following equations for the
annihilation (and creation, by taking Hermitian conjugates) operators of 
the electric field [see Appendix \ref{quadraturerelations} for notations 
and definitions]:
\beq
\label{origtwoport}
j_{\pm}(\Omega)=\widetilde{\rho}'\, k_{\pm}(\Omega)+\widetilde{\tau}\,a_{\pm}(\Omega)\,, \quad \quad 
b_{\pm}(\Omega)=\widetilde{\tau}'\,k_{\pm}(\Omega)+\widetilde{\rho}\, a_{\pm}(\Omega)\,.
\eeq
Among these four complex coefficients, $\widetilde{\rho}'$, 
the effective reflectivity from
inside the arms, determines the (free) optical resonant frequency $\omega_0+\widetilde{\Omega}$
of the system through the relation: 
\beq
\label{optres}
\widetilde{\rho}'\, e^{2 i \widetilde{\Omega} L/c}=1\,.
\eeq
[Note that the carrier frequency $\omega_0$ is assumed to be on resonance in the arm
cavity, i.e.\ $\omega_0 L/\pi c=\mathrm{integer}$.] 
It turns out that if we keep fixed the arm-cavity circulating power $I_c$, 
the mirror mass $m$ and the arm-cavity length $L$,   
the input--output relation $(\widetilde{a} - \widetilde{b})$ of the two-port system 
(\ref{origtwoport}) is completely determined by  
$\widetilde{\rho}'$ alone or equivalently by the (complex) free optical 
resonant frequency $\widetilde{\Omega}$. To show this, we first 
redefine the ingoing and outgoing dark-port fields as:
\beq
\label{redefineab}
\widetilde{a}_{\pm}(\Omega) = \frac{\widetilde{\tau}}{|\widetilde{\tau}|}\,{a}_{\pm}(\Omega)\,,\quad\quad
\widetilde{b}_{\pm}(\Omega) = \frac{\widetilde{\tau}^*}{|\widetilde{\tau}|}\,{b}_{\pm}(\Omega)\,.
\eeq
This redefinition is always possible since we can freely choose another (common) reference
point for the input and output fields. 
Secondly, using the Stokes relations given in the Appendix~\ref{stokes}, 
we derive the following equations:
\bea
\label{efftwoport}
j_{\pm}(\Omega)&=&\widetilde{\rho}' \,k_{\pm}(\Omega)+|\widetilde{\tau}|\,
\widetilde{a}_{\pm}(\Omega)=\widetilde{\rho}'\, k_{\pm}(\Omega)+\sqrt{1-|\widetilde{\rho}'|^2}\,
\widetilde{a}_{\pm}(\Omega)\,, \\
\widetilde{b}_{\pm}(\Omega)&=&|\widetilde{\tau}|\,k_{\pm}(\Omega)-
\widetilde{\rho}'^*\, \widetilde{a}_{\pm}(\Omega)
=\sqrt{1-|\widetilde{\rho}'|^2}\,k_{\pm}(\Omega)-
\widetilde{\rho}'^*\, \widetilde{a}_{\pm}(\Omega)\,,
\label{efftwoport1}
\eea
{}from which we infer that the output fields $\widetilde{b}_{\pm}(\Omega)$ depend only 
on $\widetilde{\rho}'$ or equivalently on $\widetilde{\Omega}$. 
Thus, if we vary the interferometer characteristic parameters $T$, $\rho$ and $\phi$ such that  
$\widetilde{\rho}'$ is preserved, the input--output relation do not change. 
We refer to the transformation among the interferometer parameters having 
this property as the \emph{scaling law}.
\begin{figure}[t]
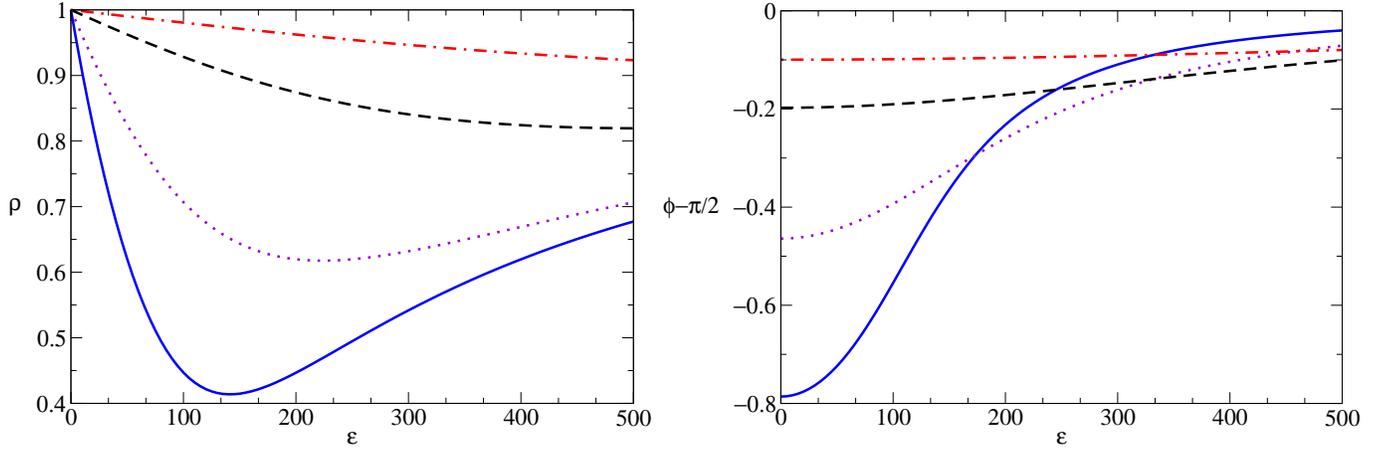

\begin{center}
\begin{tabular}{cc}
\epsfig{file=Fig3a.eps,height=0.25\textheight} & 
\epsfig{file=Fig3b.eps,height=0.25\textheight}
\end{tabular}
\caption{\label{rhophi} We plot $\rho$ and $\phi-\pi/2$ 
versus $\epsilon$ for  $\lambda =2\pi \times 100\,{\rm Hz}$ (solid line), 
$2\pi \times 200\,{\rm Hz}$ (dotted line),
$2\pi\times 500\,{\rm Hz}$ (dashed line) and $2\pi\times 1000\,{\rm Hz}$ (dashed-dotted 
line), having fixed $T=0.033$.}
\end{center}
\end{figure}
\subsection{The scaling law in terms of interferometer parameters}
\label{subsec2.3}

\begin{figure}[t]
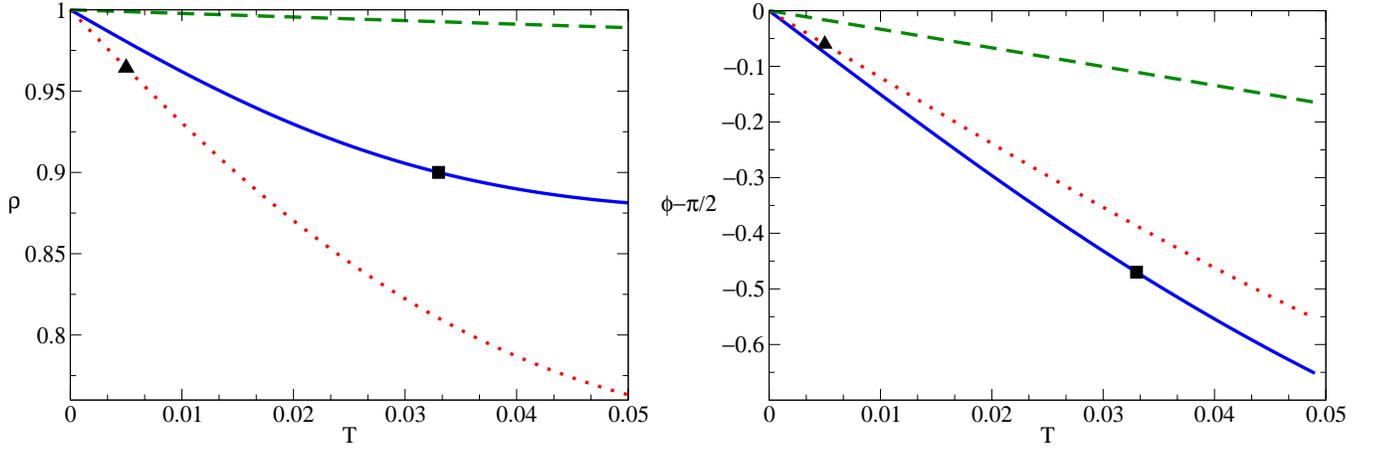

\begin{center}
\begin{tabular}{cc}
\epsfig{file=Fig4a.eps,height=0.25\textheight} & 
\epsfig{file=Fig4b.eps,height=0.25\textheight}
\end{tabular}
\caption{We plot $\rho$ and $\phi-\pi/2$ versus $T$ 
for three sets of optical resonances: 
$(\lambda,\epsilon)=(2\pi\times194.48\,{\rm
Hz},2\pi\times25.42\,{\rm Hz})$ (solid lines),
$(\lambda,\epsilon)=(2\pi\times 228.10\,{\rm Hz},2\pi\times
69.13\,{\rm Hz})$ (dotted lines) and
$(\lambda,\epsilon)=(2\pi\times 900\,{\rm Hz}, 2\pi\times30\,{\rm
Hz})$ (dashed-dotted lines). We mark with a 
square and a triangle the special configurations selected 
in Refs.~\cite{BC1,BC2,BC3}, with
$(T,\rho,\phi)=(0.033,0.9,\pi/2-0.47)$, and the current LIGO-II 
reference design~\cite{LIGOIIref}, with $(T,\rho,\phi)=(0.005,0.96,\pi/2-0.06)$, respectively. 
\label{rhophiscaling}}
\end{center}
\end{figure}

In this section we give the explicit expression of the scaling law in terms of the 
practical parameters of the SR interferometer.
We start by deriving the effective transmissivities and reflectivities 
$\widetilde{\rho}$, $\widetilde{\tau}$, $\widetilde{\rho}'$ and $\widetilde{\tau}'$ 
in terms of $T$, $R=1-T$, $\rho$ and $\phi$. 
By imposing transmission and reflection conditions at the ITM and SRM, and
propagating the fields between these mirrors (see Fig.~\ref{ITSRM}), we get the following 
equations:
\beq
\label{xy}
\tau\,\widetilde{a}_{\pm}(\Omega) + \rho\,e^{i\phi}\,y_\pm(\Omega)= x_\pm(\Omega) \,,
\quad \quad \sqrt{T}\,{k}_{\pm}(\Omega) - \sqrt{R}\,e^{i\phi}\,x_\pm(\Omega)= y_\pm(\Omega) \,,
\eeq
\beq
\label{ab}
-\rho\,\widetilde{a}_{\pm}(\Omega) + \tau\,e^{i\phi}\,y_\pm(\Omega)= \widetilde{b}_\pm(\Omega) \,,
\quad \quad \sqrt{R}\,{k}_{\pm}(\Omega) + \sqrt{T}\,e^{i\phi}\,x_\pm(\Omega)= j_\pm(\Omega) \,,
\eeq
where the reflection and transmission coefficients of ITM and 
SRM are chosen to be real, with signs $\{+\sqrt{T},-\sqrt{R}\}$, 
$\{+\tau,-\rho\}$ for light that impinges on a mirror from outside the
SR cavity; 
and $\{+\sqrt{T},+\sqrt{R}\}$, 
$\{+\tau,+\rho\}$ for light that impinges on a mirror from inside the
SR cavity.
Solving Eq.~(\ref{xy}) for $x_\pm$ and $y_\pm$ in terms  of $\widetilde{a}_\pm$ and 
$\widetilde{b}_\pm$, plugging these expressions into Eq.~(\ref{ab}) 
and comparing with Eq.~(\ref{origtwoport}) we obtain:
\beq 
\label{exactrhop}
\widetilde{\rho}'=\frac{\sqrt{R}+\rho \, e^{2
i\phi}}{1+\sqrt{R}\, \rho\, e^{2 i\phi}}\,, \quad\quad
\widetilde{\rho}=-\frac{\rho+\sqrt{R}\,e^{2 i\phi}}{1+\sqrt{R}\,
\rho\, e^{2 i\phi}}\,, \quad \quad \widetilde{\tau}'=\widetilde{\tau}=\frac{\tau \sqrt{T} e^{i
\phi}}{1+\sqrt{R} \,\rho\, e^{2 i\phi}}\,.  
\eeq 
It can be easily verified that these coefficients satisfy the Stokes relations 
(\ref{stokeap1})--(\ref{stokeap5}). The scaling law can be obtained by imposing 
that $\widetilde{\rho}'$ does not vary. This gives:
\beq
\label{scalingexact}
\frac{\sqrt{R}+\rho\,
  e^{2i\phi}}{1+\sqrt{R}\,\rho\, e^{2 i\phi}}=\mathrm{const.}\,.
\eeq
Using Eq.~(\ref{optres}), we derive the (complex) free optical resonant frequency in terms of $T$, $\rho$ and
$\phi$:
\beq
\label{exactomega}
\widetilde{\Omega}=\frac{i c}{2 L}\log\frac{\sqrt{R}+\rho
  \,e^{2i\phi}}{1+\sqrt{R}\,\rho\, e^{2 i\phi}}\equiv -\lambda -i\,\epsilon\,,
\eeq
where we trade $\widetilde{\Omega}$ for two real numbers, the 
resonant frequency $\lambda$ and decay rate (inverse decay time) $\epsilon$. 
For any choice of $T$, the parameters $\rho$ and $\phi$ 
can be expressed in terms of $\lambda$ and $\epsilon$ 
by solving Eq.~(\ref{exactomega}) in terms of $\rho\,e^{2i\phi}$. The result is:
\beq
\rho\, e^{2i\phi}=\frac{e^{-2\epsilon L/c}e^{2i\lambda
  L/c}-\sqrt{R}}{1-\sqrt{R}e^{-2\epsilon L/c} e^{2i\lambda L/c}}\,.
\label{scex}
\eeq
In  Fig.~\ref{rhophi} we plot $\rho$ (left panel) and $\phi-\pi/2$ (right panel) 
as functions of $\epsilon$ for four typical values of  
$\lambda$: $2\pi \times 100\,{\rm Hz}$ (solid lines),
$2\pi \times 200\,{\rm Hz}$ (dotted lines),
$2\pi \times 500\,{\rm Hz}$ (dashed lines) and 
$2\pi \times 1000\,{\rm Hz}$ (dashed-dotted lines), 
while fixing $T=0.033$. In Fig.~\ref{rhophiscaling}, we plot $\rho$ and
$\phi-\pi/2$ as functions of $T$, as obtained from Eq.~(\ref{scex}), 
for three sets of optical resonances: $(\lambda,\epsilon)=(2\pi\times 194.5\,{\rm 
Hz},2\pi\times 25.4\,{\rm Hz})$, plotted in solid lines, which goes through
the point $(T,\rho,\phi)=(0.033,0.9,\pi/2-0.47)$ (marked by a square), 
which is the configuration selected in Refs.~\cite{BC1,BC2,BC3};
$(\lambda,\epsilon)=(2\pi \times 228.1\,{\rm Hz},2\pi \times
69.1\,{\rm Hz})$, plotted in dotted lines, which goes through the
point $(T,\rho,\phi)=(0.005,0.96,\pi/2-0.06)$ (marked by a triangle), 
which is the current LIGO-II reference design~\cite{LIGOIIref};
and $(\lambda,\epsilon)=(2\pi \times 900\,{\rm Hz}, 2\pi \times 30\,{\rm
Hz})$, plotted in dashed-dotted lines, which is an example of a configuration
with narrowband sensitivity around a high frequency. 
As $T$, $\rho$ and $\phi$ vary along these curves, the input-output relation is preserved. 

As done in Refs.~\cite{BC1,BC2}, we now expand 
all the quantities in $T$ and keep only the first nontrivial order. 
[The accuracy of this procedure will be discussed in Sec.~\ref{sec5}.] 
For the crucial quantity $\widetilde{\rho}'$ a straightforward 
calculations gives: 
\beq
\widetilde{\rho}'=1-\frac{T}{2}\,\frac{1-\rho\,e^{2 i
\phi}}{1+\rho\,e^{2 i \phi}}\,.
\label{rhop}
\eeq
So the scaling law at linear leading order in $T$ is:
\beq
\label{sc}
T\,\frac{1-\rho\,e^{2 i \phi}}{1+\rho\,e^{2 i \phi}}=\mathrm{const.}\,.
\eeq
Moreover, applying Eq.~(\ref{rhop}) to Eq.~(\ref{optres}), 
we derive the following expression for the (free) optical resonant
frequency at leading order in $T$:
\beq
\label{FORF}
\widetilde{\Omega}
=\frac{1}{i}\,\frac{1-\rho\,e^{2 i\phi}}{1+\rho\,e^{2 i\phi}}\,\frac{Tc}{4L} 
=\frac{-2\rho\sin2\phi-i(1-\rho^2)}{1+\rho^2+2\rho\cos2\phi}\,\gamma\,, 
\eeq
where $\gamma={Tc}/{4L}$ is the half-bandwidth of the arm cavity. 
The frequency $\widetilde{\Omega}$ coincides 
with the frequency $\Omega_{-}$ introduced in Ref.~\cite{BC3}.  
[Since the authors of Ref.~\cite{BC3} used the quadrature formalism, 
they had to introduce another (free) optical resonant frequency 
which they denoted by $\Omega_{+}=-\Omega_{-}^*$. 
See discussion around Eq.~(\ref{ares}) in
Appendix~\ref{quadraturerelations}.] Thus, at linear order in $T$ we have:
\beq
\label{lambdaepsilon}
\lambda=\frac{2\rho\,\gamma\,\sin2\phi}{1+\rho^2+2\rho\cos2\phi}\,,
\quad \quad \epsilon=\frac{(1-\rho^2)\,\gamma}{1+\rho^2+2\rho\cos2\phi}\,.
\eeq
Finally, using Eqs.~(\ref{stokeap1}) and Eq.~(\ref{rhop}) we obtain the coefficients 
redefining the fields $a_\pm(\Omega)$ and $b_\pm(\Omega)$ in Eq.~(\ref{redefineab}): 
\beq
\label{argtau}
\frac{\widetilde{\tau}}{|\widetilde{\tau}|}=
\frac{(1+\rho)\cos\phi+i(1-\rho)\sin\phi}{\sqrt{1+2\rho\cos2\phi+\rho^2}}\,.
\eeq
\section{Input--output relation and noise spectral density in terms of characteristic parameters}
\label{sec3}

\subsection{Input--output relation}
\label{sec3.1}

In this section we shall express the input--output relation of SR interferometer  
(at leading order in $T$) \emph{only} in terms of the (free) optical resonant frequency, 
$\widetilde{\Omega}=-\lambda-i\,\epsilon$, and the parameter $\iota_c$, defined by
\beq
\iota_c = \frac{8\omega_0\,I_c}{m\, L\, c}\,,
\eeq
where the circulating power $I_c$ is related to the input power at BS $I_0$ by: 
\beq
\label{ici0}
I_c=\frac{2}{T}I_0\,.
\eeq
Using Eq.~(\ref{argtau}) and the results derived in Appendix~\ref{quadraturerelations} 
[see Eqs.~(\ref{bpm-apm}), (\ref{FPsi}) and  (\ref{quadraturespecial})] 
we transform Eqs.~(\ref{redefineab}), which are given in terms of annihilation and 
creation operators, into equations for quadrature fields:
\beq
\left(
\begin{array}{c}
\widetilde{a}_1 \\
\widetilde{a}_2
\end{array}
\right)
=
\frac{1}{\sqrt{1+2\rho\cos2\phi+\rho^2}}
\left(
\begin{array}{cc}
(1+\rho)\cos\phi & -(1-\rho)\sin\phi \\
(1-\rho)\sin\phi & (1+\rho)\cos\phi
\end{array}
\right)
\left(
\begin{array}{c}
a_1 \\
a_2
\end{array}
\right)\,,
\label{t1}
\eeq
and
\beq
\label{t2}
\left(
\begin{array}{c}
\widetilde{b}_1 \\
\widetilde{b}_2
\end{array}
\right)
=
\frac{1}{\sqrt{1+2\rho\cos2\phi+\rho^2}}
\left(
\begin{array}{cc}
(1+\rho)\cos\phi & (1-\rho)\sin\phi \\
-(1-\rho)\sin\phi & (1+\rho)\cos\phi
\end{array}
\right)
\left(
\begin{array}{c}
b_1 \\
b_2
\end{array}
\right)\,.
\eeq
Inserting the above expressions into Eqs.~(2.20)--(2.24) of Ref.~\cite{BC1}, and using
Eqs.~(\ref{lambdaepsilon})--(\ref{ici0}), we get the input--output relation 
depending only on the characteristic or scaling invariant quantities 
$\lambda$, $\epsilon$ and $\iota_c$: 
\beq
\label{inoutsi}
\left(
\begin{array}{c}
\widetilde{b}_1 \\
\widetilde{b}_2
\end{array}
\right)
=\frac{1}{\widetilde{M}^{(1)}}\,
\left\{
\left(
\begin{array}{cc}
\widetilde{C}^{(1)}_{11} &\widetilde{C}^{(1)}_{12} \\
\widetilde{C}^{(1)}_{21} &\widetilde{C}^{(1)}_{22} 
\end{array}
\right)
\left(
\begin{array}{c}
\widetilde{a}_1  \\
\widetilde{a}_2
\end{array}
\right)
+
\left(
\begin{array}{c}
\widetilde{D}^{(1)}_{1} \\
\widetilde{D}^{(1)}_{2}
\end{array}
\right)
\frac{h}{h_{\rm SQL}}
\right\}\,,
\eeq
where we define: 
\beq
\widetilde{M}^{(1)}={\left[\lambda^2-(\Omega+i\epsilon)^2\right]\,\Omega^2-\lambda\,\iota_c}\,,
\label{denm}
\eeq
and
\beq
\widetilde{C}^{(1)}_{11} = \widetilde{C}^{(1)}_{22}= \Omega^2(\Omega^2-\lambda^2+\epsilon^2)+ \lambda\,\iota_c \,,
\quad \quad \widetilde{C}^{(1)}_{12} = -2 \epsilon\,\lambda\, \Omega^2\,, 
\quad \quad \widetilde{C}^{(1)}_{21} = 2\epsilon\,\lambda\, \Omega^2 - 2 \epsilon\, \iota_c\,,
\label{coeffc}
\eeq
\beq
\widetilde{D}^{(1)}_{1} = -2\lambda\,\sqrt{\epsilon\, \iota_c}\,\Omega\,, 
\quad \quad \widetilde{D}^{(1)}_{2} = 2(\epsilon-i\Omega)\,\Omega\,\sqrt{\epsilon\, \iota_c}\,,
\label{coeffd}
\eeq
and 
\beq
\label{sql}
h_{\rm SQL}\equiv \sqrt{\frac{8 \hbar}{m \Omega^2 L^2}}\,,
\eeq
is the free-mass SQL for the gravitational strain $h(\Omega)$ in LIGO detectors~\cite{BK92}.
The quantity $\iota_c$ has the dimension of a frequency to the third power ($\Omega^3$). Since 
it is proportional to the laser power circulating in the arm cavity, it provides a measure 
of radiation-pressure strength. In order that radiation pressure influence interferometer 
dynamics in the frequency range interesting for LIGO, we need 
\beq
\iota_c\,\stackrel{>}{_\sim}\,\Omega_{\rm GW}^3 \quad \quad \Rightarrow \quad \quad I_c\stackrel{>}{_\sim} \frac{m\,L\,c\,\Omega^3_{\rm GW}}{8\omega_0}\,,
\label{iotac}
\eeq
which gives $I_c\stackrel{>}{_\sim}100\,$ kW for typical LIGO-II--parameters and
$\Omega_{\rm GW}=2\pi\times 100\,$Hz. 
The input--output relation (\ref{inoutsi}) is more explicit in representing 
interferometer properties than that given in Ref.~\cite{BC1}, and can be 
quite useful in the process of optimizing the SR optical 
configuration~\cite{BCM}. From the last term of Eq.~(\ref{inoutsi}) we observe that as long as 
the SR oscillation frequency $\lambda \neq 0$, both quadrature fields contain the GW signal. 
Moreover, the resonant structure, discussed in Ref.~\cite{BC1,BC2}, 
is readily displayed in the denominator of Eq.~(\ref{inoutsi}), given by Eq.~(\ref{denm}). 
As we shall see in Sec.~\ref{sec4}, the shot noise and radiation-pressure noise, 
and the fact they are correlated, can also be easily worked out from Eq.~(\ref{inoutsi}).

In Ref.~\cite{BC3} we found that one of the SR resonant frequencies, obtained by imposing 
$\widetilde{M}^{(1)}=0$, has always a positive imaginary part, corresponding to an instability.  
This instability has an origin similar to the dynamical
instability induced in a detuned Fabry-Perot cavity by
the radiation-pressure force acting on the mirrors~\cite{All,OB}.
To suppress it, we proposed~\cite{BC3} a
feed-back control system that does not compromise the GW interferometer 
sensitivity. Although the model we used to describe the servo system
may be realistic for an all-optical control loop, this might not be the case 
if an electronic servo system is implemented. However, results obtained in 
Refs.~\cite{french} would suggest it does. In any case, a more 
thorough studying should be pursued to fully clarify 
this issue. In this paper, we always assume that an appropriate control system 
of the kind proposed in Ref.~\cite{BC3} is used. 

Finally, when $\lambda=0$ (which corresponds to either $\rho=0$, or
$\rho \neq 0$, $\phi=0,\,\pi/2$) Eq.~(\ref{inoutsi}) simplifies to
\beq
\label{inouttuned}
\left(
\begin{array}{c}
\widetilde{b}_1 \\
\widetilde{b}_2
\end{array}
\right)
=
e^{2 i \beta'}
\left(
\begin{array}{cc}
1 &  0\\
-{\cal K}' & 1
\end{array}
\right)
\left(
\begin{array}{c}
\widetilde{a}_1 \\
\widetilde{a}_2
\end{array}
\right)
+e^{i\beta'}
\sqrt{2 {\cal K}'}
\left(
\begin{array}{c}
0 \\
1
\end{array}
\right)
\frac{h}{h_{\rm SQL}}\,,
\label{e1}
\eeq
which exactly coincides with Eq.~(16) of Ref.~\cite{KLMTV00} for a conventional
interferometer, but where 
\beq
\beta'=\arctan\left(\frac{\Omega}{\epsilon}\right)\,, \quad \quad 
{\cal K'}=\frac{2\,\epsilon\,\iota_c}{\Omega^2\,(\Omega^2+\epsilon^2)}\,.
\label{e2}
\eeq
The simple relations (\ref{e1}), (\ref{e2}) nicely 
unify the SR optical configuration $\phi=0, \pi/2$ (denoted by ESR/ERSE in Ref.~\cite{BC2}) 
with the conventional-interferometer optical configuration.

\subsection{Noise spectral density}
\label{sec3.2}

The noise spectral density can be calculated as follows~\cite{KLMTV00,BC1}.
Assuming that the quadrature $\widetilde{b}_\zeta =\widetilde{b}_1\,\sin \zeta +
\widetilde{b}_2\,\cos \zeta $ is measured, and using 
Eq.~(\ref{inoutsi}), we can express the interferometer noise 
as an equivalent GW Fourier component:
\beq
h_n \equiv h_{\rm SQL}\,\Delta \widetilde{b}_{\zeta}\,,
\eeq
where
\beq
\Delta \widetilde{b}_{\zeta} = \frac{(\widetilde{C}^{(1)}_{11}\,\sin\zeta+\widetilde{C}^{(1)}_{21}\,\cos\zeta)\,
\widetilde{a}_1+
(\widetilde{C}^{(1)}_{12}\,\sin\zeta+\widetilde{C}^{(1)}_{22}\,\cos\zeta)\,\widetilde{a}_2}
{\widetilde{D}^{(1)}_1\,\sin\zeta+\widetilde{D}^{(1)}_2\,\cos\zeta}\,.
\eeq
Then the (single-sided) spectral density $S^{\zeta}_h(f)$, with $f = \Omega/2\pi$, 
associated with the noise $h_n$ can be computed by the formula [see Eq.~(22) 
of Ref. \cite{KLMTV00}]:
\beq
\label{33}
2\pi\,\delta(\Omega - \Omega^\prime)\,S^\zeta_h(f) 
= \langle {\rm in} | h_n(\Omega)\,h_n^\dagger(\Omega^\prime) + 
h_n^\dagger(\Omega^\prime)\,h_n(\Omega)|{\rm in} \rangle \,.
\eeq
Assuming that the input of the whole SR interferometer 
is in its vacuum state, i.e.\ $|{\rm in} \rangle = |0_{\tilde{a}} \rangle$, and using  
\beq
 \langle 0_{\tilde{a}}| \widetilde{a}_i(\Omega)\,\widetilde{a}^\dagger_{j}(\Omega') 
+ \widetilde{a}^\dagger_{j}(\Omega') \,\widetilde{a}_i(\Omega)
|0_{\tilde{a}} \rangle= 2\pi\,\delta(\Omega- \Omega^\prime)\,
\delta_{i j}\,,
\eeq 
we find that Eq.~(\ref{33}) can be recast in the simple form (note that $\widetilde{C}^{(1)}_{ij}\in\Re$):
\beq
\label{nsd}
S_h^{\zeta}= {h_{\rm SQL}^2}\,
\frac{\left(\widetilde{C}^{(1)}_{11}\,\sin\zeta+\widetilde{C}^{(1)}_{21}\,\cos\zeta\right)^2+
\left(\widetilde{C}^{(1)}_{12}\,\sin\zeta+\widetilde{C}^{(1)}_{22}\,\cos\zeta\right)^2}
{\left|\widetilde{D}^{(1)}_1\,\sin\zeta+\widetilde{D}^{(1)}_2\,\cos\zeta\right|^2}\,.
\eeq
Plugging into the above expression Eqs.~(\ref{coeffc}), (\ref{coeffd}) we get the very explicit 
(and very simple!) expression for the noise spectral density:
\bea
\label{sdlead}
S_h^{\zeta}&=&\frac{\Omega^2 h_{\rm SQL}^2}{4 \epsilon \iota_c
  \left[\Omega^2\cos^2\zeta +(\epsilon
    \cos\zeta-\lambda\sin\zeta)^2\right]} 
\Bigg\{
\left[(\Omega+\lambda)^2+\epsilon^2\right]
\left[(\Omega-\lambda)^2+\epsilon^2\right]
+
\frac{2\iota_c}{\Omega^2}
\big[
\Omega^2(\lambda-\epsilon \sin2\zeta) \nonumber \\
&-&\lambda(\epsilon^2+\lambda^2+2\epsilon^2\cos2\zeta)
-\epsilon(\epsilon^2-\lambda^2)\sin2\zeta\big]
+
\frac{\iota_c^2}{\Omega^4}
\left[2\epsilon^2(1+\cos2\zeta)-2\epsilon\lambda\sin2\zeta+\lambda^2\right]
\Bigg\}\,.
\eea

\section{Optomechanical dynamics in terms of characteristic parameters}
\label{sec4}

The scaling laws (\ref{scex}), (\ref{sc}) could have been equivalently derived by imposing 
the invariance of the optomechanical dynamics~\cite{BC3}. In this section we express all the relevant 
quantities characterizing the SR optomechanical dynamics in terms of the scaling invariant 
parameters $\lambda$, $\epsilon$ and $\iota_c$. 

\subsection{Radiation-pressure force}
\label{4.1}

In Ref.~\cite{BC2} we assumed that SR interferometers can be
artificially divided into two linearly coupled, but otherwise
independent subsystems: the probe ${\cal P}$, which is subject to the
external classical GW force $G$ and the detector ${\cal D}$, which yields a 
classical output $Z$. The Hamiltonian of the overall system is given by 
[see Sec.~IIB in Ref.~\cite{BC2} for notations and definitions]:
\beq
H = H_{\cal P}+ {H}_{\cal D} - {x}\,({F}+ G)\,,
\eeq
where $x$ is the operator describing the antisymmetric mode of motion 
of four arm-cavity mirrors and $F$ is the radiation-pressure or back-action 
force the detector applies on the probe.
In the Heisenberg picture, using the superscript ${(1)}$ for 
operators evolving under the total Hamiltonian
${H}$, and superscript ${(0)}$ for operators evolving 
under the free Hamiltonian of the detector ${H}_{\cal D}$, the equations of motion 
in Fourier domain read~\cite{BC2}:
\bea
\label{Z1}
{Z}^{(1)}(\Omega)&=&{Z}^{(0)}(\Omega)+R_{ZF}(\Omega)\,{x}^{(1)}(\Omega)\,, \\
{F}^{(1)}(\Omega)&=&{F}^{(0)}(\Omega)+R_{FF}(\Omega)\,{x}^{(1)}(\Omega)\,, \\
{x}^{(1)}(\Omega)&=& L\,h(\Omega) + R_{xx}(\Omega)\,{F}^{(1)}(\Omega)\,.
\label{X1}
\eea
where $R_{xx}(\Omega) =-4/m/\Omega^2$, $h(\Omega)$ is the gravitational strain 
[see Eq.~(2.15) of Ref.~\cite{BC3}] related to the GW force in Fourier domain 
by $G(\Omega) = -(m/4)\,L\,\Omega^2\,h(\Omega)$, while the various 
Fourier-domain susceptibilities are defined by:
\beq
R_{AB}(\Omega)\equiv\frac{i}{\hbar}\int_{0}^{+\infty}d\tau\,e^{i\Omega\tau}\,
[A(0),B(-\tau)]\,,
\eeq
where $[A(t),B(t')]$ is the commutator between operators $A$ and $B$.
As discussed in Sec.~\ref{sec1}, LIGO-II has been planned to work at a laser power for 
which shot noise and radiation-pressure noise are comparable in the observational 
band $40$--$200$\,Hz. In Sec.~IIIA of Ref.~\cite{BC2} the radiation-pressure force 
was explicitly derived. Here, we want to express it, and the other crucial quantities 
entering the equations of motion (\ref{Z1})--(\ref{X1}) in terms of the characteristic 
parameters $\lambda$, $\epsilon$ and 
\beq
{\cal I}_c = m\,\iota_c = \frac{8\omega_0\,I_c}{L\, c}\,. 
\eeq
Using Eqs.~(\ref{lambdaepsilon}) a straightforward calculation gives the rather simple expressions:
\beq
F^{(0)}(\Omega) = \sqrt{\frac{\epsilon {\cal I}_c\,\hbar}{2}}\,
\frac{(i\Omega-\epsilon)\,\widetilde{a}_1(\Omega)+\lambda\,\widetilde{a}_2(\Omega)}{
(\Omega-\lambda + i\epsilon)\,(\Omega+\lambda+i\epsilon)}\,, 
\label{rpf}
\eeq
\bea
Z_1^{(0)}(\Omega) &=& \frac{(\lambda^2-\epsilon^2 -\Omega^2)\,\widetilde{a}_1(\Omega)+
2\lambda\,\epsilon\,\widetilde{a}_2(\Omega)}
{(\Omega-\lambda + i\epsilon)\,(\Omega+\lambda+i\epsilon)}\,,\\
Z_2^{(0)}(\Omega) &=& \frac{-2\lambda\,\epsilon\,\widetilde{a}_1(\Omega) +
(\lambda^2-\epsilon^2 -\Omega^2)\,\widetilde{a}_2(\Omega)}
{(\Omega-\lambda + i\epsilon)\,(\Omega+\lambda+i\epsilon)}\,,
\eea
\bea
R_{Z_1 F}(\Omega) &=& \sqrt{\frac{\epsilon{\cal I}_c}{2\hbar}}\,
\frac{\lambda}{(\Omega-\lambda + i\epsilon)\,
(\Omega+\lambda+i\epsilon)}\,,\\
R_{Z_2 F}(\Omega) &=& -\sqrt{\frac{\epsilon{\cal I}_c}{2\hbar}}\,
\frac{(\epsilon-i\Omega)}
{(\Omega-\lambda + i\epsilon)\,(\Omega+\lambda+i\epsilon)}\,.
\eea
The optical pumping field in detuned Fabry-Perot resonator converts the free test mass 
into an optical spring having very low intrinsic noise~\cite{OB}. The 
ponderomotive rigidity $K_{\rm pond.\,rig.}$, which 
characterizes the optomechanical dynamics in SR interferometers, is also  
responsible of the beating of the free mass SQL [see Sec.~IIIC of Ref.~\cite{BC3}] and  
its explicit expression is given by:
\beq
K_{\rm pond}(\Omega) = - R_{FF}(\Omega)= -\frac{{\cal I}_c}{4}\,\frac{\lambda}
{(\Omega-\lambda + i\epsilon)\,(\Omega+\lambda+i\epsilon)}\,.
\label{rff}
\eeq
As long as the free optical resonant frequency $\lambda$ differs from zero,
$K_{\rm pond}$ is always non-vanishing. Moreover, in order to have a
(nearly) real mechanical resonant frequency at low frequency, we require $\lambda<0$ 
[as can be obtained by imposing $K_{\rm pond}(\Omega=0)>0$.]

\subsection{Equivalence between noise correlations and change of dynamics}
\label{4.2}

As derived in Ref.~\cite{BC2,BC3}, the output of SR
interferometers, when the first or second quadrature of the
outgoing dark-port field is measured, can also be written as:
\beq
{\cal O}_{i}(\Omega) = {\cal Z}_i(\Omega) + R_{x x}(\Omega)\,
\left [{\cal F}_i(\Omega) + G(\Omega) \right ]\,, 
\quad \quad i = 1,2
\label{out}
\eeq
where:
\beq
{\cal Z}_i(\Omega)=\frac{Z^{(0)}_i(\Omega)}{R_{Z_i F}(\Omega)}\,,\quad {\cal
    F}(\Omega)=F^{(0)}(\Omega)-R_{FF}(\Omega)\frac{Z_i^{(0)}(\Omega)}{R_{Z_i
    F}(\Omega)}\,,\quad\quad i=1,2\,.
\eeq
Expressing these quantities in scaling-invariant form [here the first or second quadrature 
refers to $\widetilde{b}_{1}$ or $\widetilde{b}_2$, so the $Z_{1,2}$
discussed here are related to those in Ref.~\cite{BC2} by the rotation (\ref{t2})], we get:
\bea
\label{z1}
{\cal Z}_1(\Omega) 
&=& \sqrt{\frac{2 \hbar}{\epsilon {\cal I}_c}}\,\frac{1}{\lambda}\,
\left [(\lambda^2-\epsilon^2-\Omega^2)\,\widetilde{a}_1(\Omega)
+2\epsilon\,\lambda\,\widetilde{a}_2(\Omega) \right ]\,,\\
{\cal Z}_2(\Omega) 
&=& \sqrt{\frac{2 \hbar}{\epsilon {\cal I}_c}}\,\frac{1}{(\epsilon-i\Omega)}\,
\left [2\epsilon\,\lambda\,\widetilde{a}_1(\Omega) - 
(\lambda^2-\epsilon^2-\Omega^2)\,\widetilde{a}_2(\Omega) \right ]\,,
\eea
and
\bea
{\cal F}_1(\Omega) 
&=& \sqrt{\frac{{\cal I}_c\,\hbar}{8\,\epsilon}}\,\widetilde{a}_1(\Omega)
\,,\\
{\cal F}_2(\Omega) 
&=& \sqrt{\frac{{\cal I}_c\,\hbar}{8\,\epsilon}}\,\frac{1}{(\epsilon-i\Omega)}\,
\left [2\epsilon\,\widetilde{a}_1(\Omega) - \lambda\,\widetilde{a}_2(\Omega)) \right ]\,. 
\label{f2}
\eea
The form of Eq.~(\ref{out}), along with the fact that the operators
${\cal Z}_i(\Omega)$ and ${\cal F}_i(\Omega)$ are proportional  
to $1/\sqrt{I_c}$ and  $\sqrt{I_c}$, made it natural to  refer to them~\cite{BC3} as 
\emph{effective} output fluctuation and \emph{effective} radiation-pressure 
force. The quantum noise embodied in ${\cal Z}_i(\Omega)$ is the shot noise, while 
the quantum noise described by ${\cal F}_i(\Omega)$ is the radiation-pressure or 
back-action noise.
The operators ${\cal Z}_i(\Omega)$, ${\cal F}_i(\Omega)$ 
satisfy the following commutation relations~\cite{BK92,BC2,BC3}:
\beq
[{\cal Z}_i (\Omega), {\cal Z}_i^\dagger (\Omega')] = 0 = [{\cal F}_i(\Omega), 
{\cal F}_i^\dagger (\Omega')]\,,
\quad \quad [{\cal Z}_i(\Omega), 
{\cal F}_i^\dagger (\Omega')]= -2\pi\,i\,\hbar\,\delta(\Omega-\Omega')
\,, \quad \quad i = 1,2\,. 
\label{comm}
\eeq
If the output quadrature $i$ is measured, the noise spectral density (\ref{nsd}), 
written in terms of the operators ${\cal Z}_i$ and ${\cal F}_i$, reads~\cite{BK92}: 
\beq
S_{h,i}(\Omega)=\frac{1}{L^2}\,\left\{
       S_{{\cal Z}_i {\cal Z}_i}(\Omega)
      +2{\cal R}_{xx}(\Omega)\,\Re\left[S_{{\cal F}_i {\cal Z}_i}(\Omega)\right] 
+ {\cal R}_{xx}^2(\Omega)\, S_{{\cal F}_i {\cal F}_i}(\Omega)\right\}\,, 
\label{nsdo}
\eeq
where the (one-sided) cross spectral density of two operators 
is expressible, by analogy with Eq.~(\ref{33}), as
\beq
2\pi\,\delta\left(\Omega-\Omega'\right)\,S_{{\cal A} {\cal B}}(\Omega) = 
\langle 0_{\tilde{a}}| {\cal A}(\Omega) {\cal B}^\dagger(\Omega')+ {\cal B}^\dagger
(\Omega') {\cal A}(\Omega)| 0_{\tilde{a}}\rangle\,.
\eeq
In Eq.~(\ref{nsdo}), the terms containing $S_{{\cal Z}_i {\cal Z}_i}$, $S_{{\cal F}_i 
{\cal F}_i}$ and $\Re\left[S_{{\cal F}_i {\cal Z}_i}\right]$ 
should be identified as shot noise, radiation-pressure noise and a term 
proportional to the correlation between the two noises,
respectively \cite{BK92}. The noise spectral densities expressed in terms of 
the scaling invariant quantities $\lambda$, $\epsilon$ and ${\cal I}_c$ are rather 
simple and read: 
\bea
S_{{\cal Z}_1 {\cal Z}_1}(\Omega) &=& 
\frac{2\hbar}{{\cal I}_c}\,
\frac{\left[(\Omega+\lambda)^2+\epsilon^2\right]\left[(\Omega-\lambda)^2+\epsilon^2\right]}{\epsilon\,\lambda^2}\,,\\
S_{{\cal Z}_2 {\cal Z}_2}(\Omega) 
&=& \frac{2\hbar}{{\cal I}_c}\,\frac{
\left[(\Omega+\lambda)^2+\epsilon^2\right]\left[(\Omega-\lambda)^2+\epsilon^2\right]
}{\epsilon\,(\epsilon^2+\Omega^2)}\,,
\label{Z2Z2}
\eea
\bea
S_{{\cal F}_1 {\cal F}_1}(\Omega) &=& \frac{\hbar{\cal I}_c}{8\,\epsilon}\,,\\
S_{{\cal F}_2 {\cal F}_2}(\Omega) &=& \frac{\hbar{\cal I}_c}{8\,\epsilon}\, \frac{(4\epsilon^2 + \lambda^2)}{\epsilon^2+\Omega^2}\,,
\eea
\bea
S_{{\cal Z}_1 {\cal F}_1}(\Omega) &=& \hbar\,\frac{(\lambda^2-\epsilon^2-\Omega^2)}{2\epsilon\,\lambda}\,,\\
S_{{\cal Z}_2 {\cal F}_2}(\Omega) &=& \hbar\,\frac{\lambda\,(\lambda^2+3\epsilon^2-\Omega^2)}{2\epsilon\,(\epsilon^2+\Omega^2)}\,.
\eea
Note that in our case $S_{{\cal F}_i {\cal Z}_i}$ is real, thus 
$S_{{\cal F}_i {\cal Z}_i}=S_{{\cal Z}_i {\cal F}_i}$.
It is straightforward to check that the following relation is also satisfied:
\beq
\label{Heisenberg}
S_{{\cal Z}_i {\cal Z}_i}(\Omega)\,S_{{\cal F}_i {\cal F}_i}(\Omega) - 
S_{{\cal Z}_i {\cal F}_i}(\Omega)\,S_{{\cal F}_i {\cal Z}_i}(\Omega) = 
\hbar^2\,, \quad \quad i = 1,2\,.
\eeq
Since in SR interferometers $S_{{\cal Z}_i {\cal F}_i} \neq 0$, the noise spectral density 
$S_{h,i}$ is not limited by the free-mass SQL for GW interferometers ($S_{\rm SQL} \equiv h^2_{\rm SQL}$), 
as derived and discussed in Refs.~\cite{BC1,BC2,BC3}. 

We want to show now that cross correlations between shot noise and radiation-pressure noise
are equivalent to some modification of the optomechanical dynamics of the system 
composed of probe and detector, as originally pointed out by Syrtsev and 
Khalili in Sec.~III of Ref.~\cite{SK94}. More specifically, we shall show that 
for linear quantum measurement devices, at the cost of modifying the optomechanical 
dynamics, the measurement process can be described in terms of new operators 
${\cal Z}'$ and ${\cal F}'$ with zero cross correlation. 

In Ref.~\cite{BC2} the authors found that the most generic transformation which 
preserves the commutation relations (\ref{comm}) is of the form 
[see Eq.~(2.25) in Ref.~\cite{BC2}]:
\beq
\left(
\begin{array}{c}
{\cal Z}_i'(\Omega) \\
{\cal F}_i'(\Omega)
\end{array}
\right)
=
e^{i \alpha}\,
\left(
\begin{array}{cc}
L_{11} &  L_{12}\\
L_{21} & L_{22}
\end{array}
\right)
\left(
\begin{array}{c}
{\cal Z}_i(\Omega) \\
{\cal F}_i(\Omega)
\end{array}
\right)\,,
\label{transf}
\eeq
with $\alpha,\,L_{i j} \in \Re$ and $\det L_{i j}=1$. Under this
transformation the output (\ref{out}) becomes: 
\beq
{\cal O}_i(\Omega) = e^{-i\alpha}\,\left [L_{22}-R_{xx}(\Omega)\,L_{21} \right ]\,{\cal Z}'_{i}(\Omega)
+ e^{-i\alpha}\,\left [-L_{12}+R_{xx}(\Omega)\,L_{11} \right ]\,{\cal F}'_i(\Omega) + 
R_{xx}(\Omega)\,G(\Omega)\,.
\eeq
By imposing that the system responds in the same way to electromagnetic and gravitational forces,  
${\cal F}'(\Omega)$ and $G(\Omega)$, we find the two conditions:
$e^{i\alpha}=\pm1$ and $R_{xx}(\Omega)\,(L_{11}\mp1) = L_{12}$. 
The transformation we have to apply so that the correlations 
between new fields ${\cal Z}'_i(\Omega)$  and ${\cal F}_i'(\Omega)$  are zero, 
give the following set of equations: 
\beq
L\,
\left(\begin{array}{cc} 
S_{{\cal Z}_i {\cal Z}_i}(\Omega) & S_{{\cal Z}_i {\cal F}_i}(\Omega)\\
S_{{\cal F}_i {\cal Z}_i}(\Omega) & S_{{\cal F}_i {\cal F}_i}(\Omega)
\end{array}\right)\,
L^t = 
\left(\begin{array}{cc} 
S_{{\cal Z}_i' {\cal Z}'_i}(\Omega) & 0\\
0 & S_{{\cal F}_i' {\cal F}_i'}(\Omega)
\end{array}\right)\,.
\eeq
When $S_{\cal ZF}=S_{\cal FZ}\in\Re$, as it happens in SR interferometers, 
the above conditions can be solved in infinite ways. A simple solution, 
suggested by Syrtsev and Khalili~\cite{SK94}, is obtained by 
taking $\alpha=0$ and  $L_{11}=1$. 
In this case, a straightforward calculation gives:  
$L_{12}=0$, $L_{21}=- S_{{\cal Z}_i {\cal F}_i}/S_{{\cal Z}_i {\cal Z}_i}$
and $L_{22}=1$. The output becomes: 
\beq
{\cal O}'_{i}(\Omega) = 
{\cal Z}'_{i}(\Omega) + \chi_{i}^{\rm eff}(\Omega)\, \left [{\cal F}'_{i}(\Omega) + G(\Omega) \right ]\,, 
\quad \quad 
{\cal O}'_{i}(\Omega) = {\cal O}_{i}(\Omega)\,\frac{R_{xx}(\Omega)}{\chi_{i}^{\rm eff}(\Omega)}\,, 
\eeq
where $\chi_i^{\rm eff}$, the \emph{effective} susceptibility, is given by:
\beq
\chi_{i}^{\rm eff}(\Omega) = \frac{R_{xx}(\Omega)}{1 + R_{xx}(\Omega)\,S_{{\cal Z}_i 
{\cal F}_i}(\Omega)/S_{{\cal Z}_i {\cal Z}_i}(\Omega)}\,.
\eeq
The spectral densities of the new operators ${\cal Z}_i'$ and ${\cal F}_i'$ are:
\beq
\label{ZiZi}
S_{{\cal Z}'_i {\cal Z}'_i}(\Omega) = S_{{\cal Z}_i {\cal Z}_i}(\Omega)\,, \quad \quad 
S_{{\cal F}'_i {\cal F}'_i}(\Omega) = S_{{\cal F}_i {\cal F}_i}(\Omega) - 
\frac{S_{{\cal Z}_i {\cal F}_i}^2(\Omega)}{S_{{\cal Z}_i {\cal Z}_i}(\Omega)}\,,
\quad \quad i = 1,2\,,
\eeq
with 
\bea
S_{{\cal F}'_1 {\cal F}'_1}(\Omega) &=& \frac{\hbar {\cal I}_c}{2}\,
\frac{\epsilon\,\lambda^2}{\left
  [(\Omega+\lambda)^2+\epsilon^2\right]\left[(\Omega-\lambda)^2+\epsilon^2\right]}\,,\\  
S_{{\cal F}'_2 {\cal F}'_2}(\Omega) &=& \frac{\hbar {\cal I}_c}{2}\,
\frac{\epsilon(\epsilon^2+\Omega^2)}{\left
  [(\Omega+\lambda)^2+\epsilon^2\right]\left[(\Omega-\lambda)^2+\epsilon^2\right]}\,.
\eea 
These new operators satisfy the condition [see Eq.~(\ref{Heisenberg})]:
\beq 
S_{{\cal Z}'_i {\cal Z}'_i}(\Omega)\,S_{{\cal F}'_i {\cal F}'_i}(\Omega)=\hbar^2\,, 
\quad \quad i = 1,2\,.
\label{constr}
\eeq 
\subsection{Equivalence to a single detuned cavity and frequency-dependent rigidity}
\label{4.3}

At the end of Sec.~\ref{subsec2.2} we discussed under which assumptions 
radiation-pressure effects were included in the description of SR interferometers
in Refs.~\cite{BC1,BC2,BC3}. There, the authors assumed that radiation pressure 
forces acting on ETM and ITM are equal, and disregarded ETM and ITM motions 
during the light round-trip time in arm cavities. In this case the ITM and SRM  
can be considered fixed, and as shown in Sec.~\ref{subsec2.1} it is possible to 
map the SR optical configuration to a three-mirror cavity with only the ETM
movable. We shall see explicitly in this section that, since the very short SR cavity 
can be regarded as a single effective mirror, we can further map the SR interferometer to a
single-detuned cavity with only the ETM  movable, which is exactly
the system that Khalili discussed in Ref.~\cite{FK}. [More specifically, 
the single-detuned cavity has (complex) free optical resonant
frequency $\omega_0-\lambda-i\epsilon$, ETM mass $\mu_{\rm arm}=m_{\rm 
arm}/2=m$, and circulating power $I_{\rm arm}=2I_c$. See Eqs.~(\ref{Iarm}), (\ref{marm}) and 
(\ref{muarm}).] 

If the output quadrature $i$ is measured, the noise spectral density 
expressed in terms of the operators ${\cal Z}_i'$ and ${\cal F}_i'$, 
can be written as:
\beq
S_{h,i}(\Omega) = \frac{R_{xx}^2(\Omega)}{L^2}\,\left [ \left [ \chi^{\rm eff}_i(\Omega)\right ]^{-2}\,
S_{{\cal Z}'_i {\cal Z}'_i}(\Omega) + S_{{\cal F}'_i {\cal F}'_i}(\Omega) \right ]\,.
\label{sp}
\eeq
In order to make explicit the connection with Ref.~\cite{FK}, we evaluate 
the noise spectral density for $x_{\rm GW} \equiv  L\,h/2$ and we denote it by 
$S_{x_{\rm GW}}$. It reads:
\beq
\label{SxGW}
S_{x_{\rm GW},i}(\Omega) = \frac{1}{\mu_{\rm arm}^2\,\Omega^4}\,\left \{ \left [ \frac{\chi^{\rm eff}_i(\Omega)}{4}\right ]^{-2}\,
\frac{S_{{\cal Z}'_i {\cal Z}'_i}(\Omega)}{4} + 4S_{{\cal F}'_i {\cal F}'_i}(\Omega) \right \}\,,
\eeq
where as discussed above $\mu_{\rm arm} = m_{\rm arm}/2=m$. By rewriting the generalized susceptibility into
the form,
\beq
\frac{\chi^{\rm eff}_i(\Omega)}{4} = 
\frac{1}{-\mu_{\rm arm}\,\Omega^2+4\,K^{\rm eff}_i(\Omega)}\,, \quad \quad i = 1,2\,,
\eeq
we introduce, as Khalili also did~\cite{FK}, the \emph{effective} rigidity 
$K_i^{\rm eff}(\Omega)$,  defined by:
\beq
K^{\rm eff}_i(\Omega) \equiv 
\frac{S_{{\cal Z}_i {\cal F}_i}(\Omega)}{S_{{\cal Z}_i {\cal Z}_i}(\Omega)}\,.
\eeq
More explicitly, 
\bea
\label{rig1}
K^{\rm eff}_1(\Omega) &=& \frac{{\cal I}_c \lambda}{4}\,\frac{-\epsilon^2+\lambda^2-\Omega^2}
{\left[(\Omega-\lambda)^2+\epsilon^2\right]\left[(\Omega+\lambda)^2+\epsilon^2\right]}\,,\\
\label{rig2}
K^{\rm eff}_2(\Omega) &=& \frac{{\cal I}_c
  \lambda}{4}\,\frac{3\,\epsilon^2+\lambda^2-\Omega^2}
{\left[(\Omega-\lambda)^2+\epsilon^2\right]\left[(\Omega+\lambda)^2+\epsilon^2\right]}\,.
\eea
Those expressions, in particular Eqs.~(\ref{SxGW}), (\ref{rig2}) agree with those
derived by Khalili~\cite{FK} for single detuned cavity [see Eqs.~(19) and (21) 
in Ref.~\cite{FK}] if we make the following identifications (this paper $\rightarrow$ Khalili): 
$\lambda \rightarrow \delta$, $\epsilon \rightarrow \gamma$, $2L\,I_{\rm arm}/c \equiv 4L\,I_c/c 
\rightarrow {\cal E}$ (energy stored in the single cavity),
$\chi_i^{\rm eff}/4 \rightarrow \chi$, and $4K_i^{\rm eff} \rightarrow
K$. Note that in Ref.~\cite{FK} 
it is always assumed that the second quadrature is measured.

\begin{figure}
\begin{center}
\epsfig{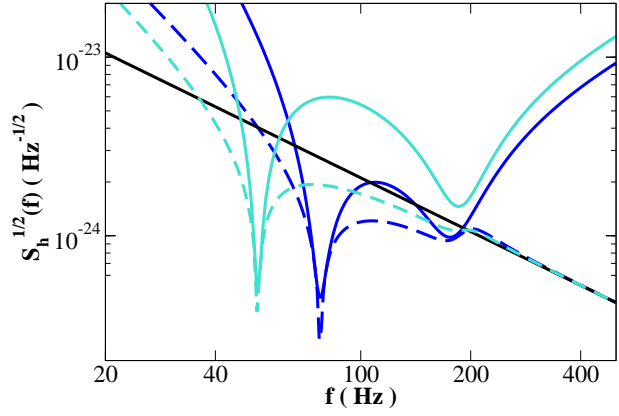}
\vspace{0.2cm}
\caption{\label{SQL} Plot of $\sqrt{S_{h,2}}$ (continuous lines) and $
\sqrt{S^{\rm min}_{h,2}}$ (dashed lines) versus frequency 
$f$ for  $T=0.033$, $\epsilon = 2 \pi \times 25.0 \,{\rm Hz}$, $\lambda = 2 \pi \times 191.3 \,{\rm Hz}$, and two different 
values of the laser power circulating in 
the arm cavities: $I_c = 300 $ kW (lighter-color lines) and $I_c = 600$ kW 
(darker-color lines). The free-mass SQL line (black straight line) is also shown for comparison.}
\end{center}
\end{figure}
The description of the measurement system in terms of the uncorrelated fields, 
${\cal Z}'_i$ and ${\cal F}'_i$, yields another way of understanding why in SR interferometers 
the free mass SQL, $S_{h}^{\rm SQL} \equiv h^2_{\rm SQL}$, loses its significance.
Indeed, by using Eq.~(\ref{constr}), we get $S_{{\cal Z}'_i {\cal Z}'_i} = 
\hbar^2/S_{{\cal F}'_i {\cal F}'_i} $. Plugging this expression into Eq.~(\ref{sp}), minimizing 
with respect to $S_{{\cal F}'_i {\cal F}'_i}$, we obtain,
\beq
S^{\rm min}_{{\cal F}'_2 {\cal F}'_2}(\Omega) = \frac{\hbar}{|1 + R_{xx}(\Omega)\,K^{\rm eff}_2(\Omega)|}\,\frac{1}{R_{xx}(\Omega)}\,,
\eeq
and the minimal noise spectral density is,
\beq
\label{min}
S_{h,i}^{\rm min}(\Omega) =
\frac{2\hbar}{L^2}\,\left|\frac{R^2_{xx}(\Omega)}{\chi^{\rm eff}_i(\Omega)}\right|=S_h^{\rm
  SQL}\left|\frac{R_{xx}(\Omega)}{\chi_i^{\rm eff}(\Omega)}\right|\,,
\eeq
which can be formally regarded as a non-free-mass SQL for the {\it effective} dynamics
described by $\chi_i^{\rm eff}$. 
To give an example, in Fig.~\ref{SQL} we plot the square root of the noise spectral densities  
$S_{h,2}$ and $S^{\rm min}_{h,2}$ versus frequency $f$ having fixed 
$\epsilon= 2 \pi \times 25.0  \,{\rm Hz}$, $\lambda = 2 \pi \times 191.3 \,{\rm Hz}$, for two different 
values of the laser power circulating in 
the arm cavities: $I_c = 300 $ kW and $I_c = 600$ kW. For comparison we also plot the 
free-mass SQL line. As we can see from the plot, $S^{\rm min}_{h,2}$
can go quite below the free-mass SQL. 

The effective dynamics can be also used to optimize the performance of 
SR interferometers~\cite{FK}. The roots of the following equation, 
\beq
K^{\rm eff}_i(\Omega) - \frac{m}{4}\,\Omega^2 = 0\,,
\label{poles}
\eeq
corresponds to resonances produced by the effective
rigidity, at which $\chi^{\rm eff}\rightarrow\infty$ and, using Eq.~(\ref{min}), 
\beq
S_{h,i}^{\rm min}(\Omega) \rightarrow 0\,.
\eeq
As observed by Khalili~\cite{FK}, we could expect that the more the roots of 
Eq.~(\ref{poles}) coincide, the more broadband the noise curve will be. 
For example, we could expect that interferometer configurations with 
double or triple zeros be optimal. However, as we shall see, those
configurations are not much better than some of the three-single-zero cases. 

Assuming the second quadrature ($i=2$) is observed, we obtain for the triple-zero case 
[see also Eqs.~(29), (30) and (31) in Ref.~\cite{FK}]:
\beq
\label{tp}
\iota_c = 2 \left ( \frac{9\sqrt{177}-113}{49} \right)\,\lambda^3\,, \quad \quad 
\epsilon = \frac{\sqrt{280-21\,\sqrt{177}}}{7}\,\lambda\,, \quad \quad 
\Omega_{\rm triple \,zero} = \sqrt{\frac{2(-11+\sqrt{177})}{7}}\,\lambda\,.
\eeq
\begin{figure}
\begin{center}
\epsfig{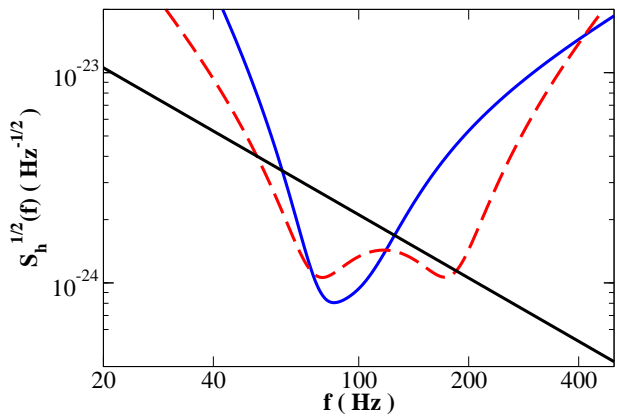}
\vspace{0.2cm}
\caption{\label{3pole} Plot of the square root of the noise spectral density $S_h$ versus frequency $f$ 
for (i) triple-zero case (continuous line) with $\lambda = 2 \pi \times 123.2\, {\rm Hz}$, 
$\epsilon = 2 \pi \times 13.8\, {\rm Hz}$, and $I_c = 320 $ kW and 
(ii) three-single-zero case (dashed line) with $\lambda = 2 \pi \times 191.3\, {\rm Hz}$,  
$\epsilon = 2 \pi \times 25.0\, {\rm Hz}$ and $I_c = 590$ kW.
For comparison we also show the free-mass SQL line (black straight line).}
\end{center}
\end{figure}
In Fig.~\ref{3pole} we plot the square root of the noise spectral density $S_{h,2}$ versus frequency $f$ 
for the triple-zero case having fixed $\Omega_{\rm triple \,zero} = 2\pi\times 100$ Hz, i.e.\ 
the (free) oscillation frequency $\lambda = 2\pi \times 123.3\, {\rm Hz}$ and $\epsilon = 2 \pi \times 13.8\, {\rm Hz}$.
The SQL line is also plotted. 
For comparison we also show the noise spectral density $S_{h,2}$ corresponding to a solution 
of Eq.~(\ref{poles}) with three-single zeros: 
$\lambda = 2 \pi \times 191.3\, {\rm Hz}$,  $\epsilon = 2 \pi \times 25.0\, {\rm Hz}$ 
and $I_c = 590$ kW. As mentioned, 
the spectral density in the triple-zero case is not significantly broad-band, especially if  
compared with the three-single-zero case. 

This result originates from the {\it non-universal} nature of the
curve $S_{h,i}^{\rm min}$. The SQL (\ref{sql}) does not change 
if we adjust (by varying the circulating power) 
the balance between shot noise and radiation-pressure noise and find 
the interferometer parameters whose noise curve can touch it. 
By contrast, the curve $S_{h,i}^{\rm min}$ changes when 
we adjust (by varying the circulating power  or the
optical resonant frequencies) the effective shot and radiation-pressure noises, 
$S_{{\cal Z}'_i {\cal Z}'_i}$ and $S_{{\cal F'}_i {\cal F}'_i}$. 
[The change of $S_{h,2}^{\rm min}$ as 
$I_c$ is varied can be also seen from Fig.~\ref{SQL}.] 
As a consequence, the fact that $S_{h,i}^{\rm min}$ is low and broad-band 
for a certain configuration cannot guarantee the noise curve will also
be optimal. In particular, in
the triple-zero case, Eq.~(\ref{poles}) already  fixes all the
interferometer parameters, leaving no freedom for the noise curve to
really take advantage of the triple zeros.  
The fact that only a {\it non-universal} minimum noise spectral
density exists in SR interferometers arises in part because of the double
role played by the carrier light. Indeed, the latter provides the means for
measurement, and therefore determines the balance between shot and
radiation-pressure noises, but it also directly affects the optomechanical 
dynamics of the system, originating the optical-spring effect. 

Finally, Braginsky, Khalili and Volikov~\cite{BKV} have recently proposed 
a table-top quantum-measurement experiment to (i) investigate the 
ponderomotive rigidity effect present in single detuned cavity 
and (ii) beat the free mass 
SQL. Although the table-top experiment will concern physical parameters very different from LIGO-II, e.g.,  
the test mass $m \sim 2 \times 10^{-2}$ g, $L \sim 1$ cm, 
$\Omega\sim10^4\,s^{-1}$, $I_c\sim 1\mbox{-}10\,{\rm W}$, however, 
because of the equivalence we have explicitly demonstrated between SR interferometers and single detuned cavities, 
the results of the table-top experiment could  shed new light and investigate various features of 
SR optomechanical configurations relevant for LIGO-II.

\subsection{Optical spring equivalent to mechanical spring but at zero temperature}
\label{4.4}

When proposing the optical-bar GW detectors~\cite{OB}, 
Braginsky, Gorodetsky and Khalili pointed out that detuned optical pumping field in 
Fabry-Perot resonator can convert the free test mass 
into an optical spring having \emph{very low} intrinsic noise. In this
section we illustrate why this happens in SR interferometers 
and why optical springs are indeed preferable to mechanical 
springs in measuring very tiny forces.

The Heisenberg operator in Fourier domain $x^{(1)}(\Omega)$ 
describing the antisymmetric mode of motion of SR interferometer, 
satisfies the following equation [see Eqs.~(\ref{Z1}), (\ref{X1}) above and also Eq.~(2.20) of Ref.~\cite{BC2}]:
\beq
x^{(1)}(\Omega) = \chi(\Omega)\,F^{(0)}(\Omega)\,, \quad 
\quad \chi(\Omega) = \frac{R_{xx}(\Omega)}{1-R_{xx}(\Omega)\,R_{FF}(\Omega)}\,.
\eeq
Using Eq.~(\ref{rff}) we get:
\beq
\label{chi}
\chi(\Omega) =  \frac{4}{m}\,\frac{\lambda^2 +(\epsilon-i\Omega)^2}
{\lambda\,\iota_c - \Omega^2\,[\lambda^2+(\epsilon-i\Omega)^2]}\,.
\eeq
\begin{figure}[t]
\begin{center}
\vspace{0.5cm}
\epsfig{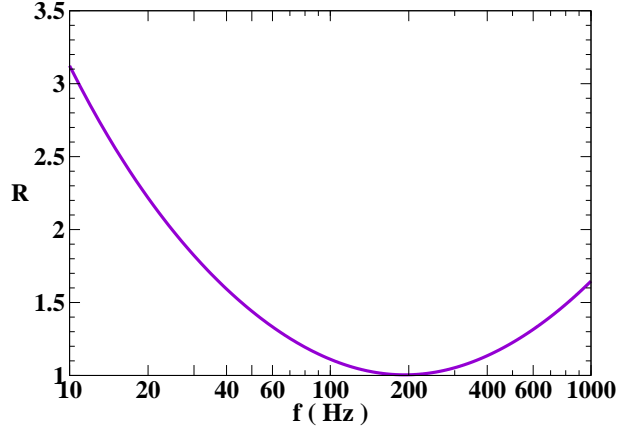}
\vspace{0.2cm}
\caption{\label{FDT} Plot of $R \equiv S_x(f)/(2\hbar\,|\Im[\chi(f)]|)$ 
versus $f$ when $\lambda = 2 \pi \times 191.3\, {\rm Hz}$,  
$\epsilon = 2 \pi \times 25.0\, {\rm Hz}$ and $I_c = 590$ kW.}
\end{center}
\end{figure}
The noise spectral density associated with $x$ is:
\beq
\label{sx}
S_{x}(\Omega) = |\chi(\Omega)|^2\,S_{F}(\Omega)\,,
\eeq
where:
\beq
\pi\,\delta\left(\Omega-\Omega'\right)\,
S_{x}(\Omega) = \langle 0_{\tilde{a}}|\, 
x^{(1)}(\Omega) x^{(1)\,\dagger}(\Omega')|0_{\tilde{a}}\rangle \,,
\quad \quad 
\pi\,\delta\left(\Omega-\Omega'\right)\,S_{F}(\Omega)= 
\langle 0_{\tilde{a}}|\,F^{(0)}(\Omega) F^{(0)\,\dagger}(\Omega')
|0_{\tilde{a}} \rangle\,. 
\eeq
More explicitly, 
\beq
\label{sf}
S_F(\Omega) = \frac{{\cal I}_c\,\hbar}{2}\,\frac{\epsilon\,(\lambda^2+\epsilon^2+\Omega^2)}
{[(\Omega-\lambda)^2+\epsilon^2]\,[(\Omega+\lambda)^2+\epsilon^2]}\,.
\eeq
For the optical spring, which is made up of electromagnetic
oscillators in their ground states (the vacuum state), we have
[see e.g., Chapter 6 in Ref.~\cite{BK92}~\footnote{~Note that the 
factor 2 in the RHS of Eqs.~(\protect\ref{qs}), (\protect\ref{ms}) 
is due to the fact that we use one-sided spectral densities 
while Braginsky and Khalili~\protect\cite{BK92} use two-sided spectral densities.}]: 
\beq
S_x(\Omega) \geq 2\hbar \, |\Im[\chi(\Omega)]|\,,
\label{qs}
\eeq
which can be regarded as a zero-temperature version of the fluctuation-dissipation
theorem. For a mechanical system, e.g., a mechanical spring, with the same susceptibility, 
but in thermal equilibrium at temperature $T\gg \hbar\Omega/k$, where $k$ is the Boltzmann
constant, the standard version of fluctuation-dissipation theorem says,
\beq
S_x(\Omega) = 4\frac{k\,T}{\Omega}\, |\Im[\chi(\Omega)]|\,.
\label{ms}
\eeq
If we assume $\Omega\sim2\pi\times100\,{\rm Hz}$,
$\hbar\Omega/k\sim 5\times 10^{-9}\,{\rm Kelvin}$, the condition $T
\gg\hbar\Omega/k$ is always valid for any practical mechanical
system. As a consequence, 
\beq
\label{dd}
S_x^{\rm mech. \,spring}(\Omega) \sim
 \frac{k\,T}{\hbar\,\Omega}\,S_x^{\rm opt.\, spring}(\Omega)\,. 
\eeq
At $T=300K$, $\Omega/2 \pi = 100$ Hz, we get 
$S_x^{\rm mech.\, spring}\sim 10^{11}\,S_x^{\rm opt.\, spring}$.
Thus, because of the very large coefficient $kT/\hbar\Omega$ in Eq.~(\ref{dd}), 
fluctuating noise in an optical spring is always much smaller than in a mechanical spring! 

For SR interferometers described in this paper, 
the fluctuating noise $S_x$ does not saturate the 
inequality in Eq.~(\ref{qs}). This can be inferred from 
Fig.~\ref{FDT} where we plot $R\equiv S_x(f)/(2\hbar\,|\Im[\chi(f)]|)$ 
versus $f$, where $S_x$ has been obtained from Eqs.~(\ref{chi}), (\ref{sx}) and 
(\ref{sf}), for the following choice of the physical parameters:
 $m=30$ kg, $T=0.033$, $\gamma=2 \pi \times 98.5 \,{\rm Hz}$,  
with $\lambda = 2 \pi \times 191.3\, {\rm Hz}$,  
$\epsilon = 2 \pi \times 25.0\, {\rm Hz}$ and $I_c = 560$ kW.
The minimum of $R$ is at the frequency corresponding to the 
(free) oscillation frequency of the SR interferometer, i.e.\ 
$f_{\rm min} = \lambda/(2 \pi) = 191.3$ Hz.

\section{Input--output relation at all orders in transmissivity of internal test-mass mirrors}
\label{sec5}

To simplify the calculation and the modeling of GW interferometers, 
KLMTV~\cite{KLMTV00} calculated  the input--output relation 
of a conventional interferometer at leading order in $T$ 
and $\Omega\,L/c$. By taking only the leading order terms in $T$, they 
ignored the radiation-pressure forces acting on the 
ITM due to the electromagnetic field present in the cavity made up  
of ITM and BS. By limiting their analysis to the leading order 
in $\Omega L/c$, they assumed the radiation-pressure forces acting on the ITM
and ETM are equal. 
In conventional interferometers, $T$ alone determines the
half-bandwidth $\gamma$ of the arm cavities (through $\gamma=T c/4
L$), which fully characterizes the interferometer 
[see Eq.~(16) in Ref.~\cite{KLMTV00} and Eqs.~(\ref{e1}), (\ref{e2}) above]. 
Moreover, since $\Omega_{\rm GW}$ is comparable to $\gamma$ and $T \sim 0.005 - 0.033$, 
the two small quantities, $\Omega L/c$ and $T$ are on the same order,
and the accuracy in expanding the input--output relation in these 
two parameters is rather under control. 
[Note that if $\gamma\sim2\pi\times100\,{\rm Hz}$, we have $T\sim0.033$.]
  
In describing SR interferometers, the authors of Refs.~\cite{BC1,BC2,BC3} 
build on the leading-order results of Ref.~\cite{KLMTV00}. 
However, in SR interferometers the accuracy of expanding in $T$ can be 
quite obscure, because $T$ is not the
only small quantity characterizing SR-interferometer performances 
--- for example the SRM transmissivity can also be a small quantity. 
Thus, to clarify the accuracy of the expansion in $T$, we now derive the 
input--output relation at all orders in $T$, and compare with the leading order result 
(\ref{inoutsi})~\cite{BC1,BC2}. The calculation is much easier if we view the SR cavity as a 
single effective mirror, as done in Sec.~\ref{sec2}.  However, in
doing so, we still use the assumptions mentioned at the beginning of
this section. See also the end of Sec.~\ref{subsec2.1}. 
\begin{figure}
\begin{center}
\begin{tabular}{cc}
\epsfig{file=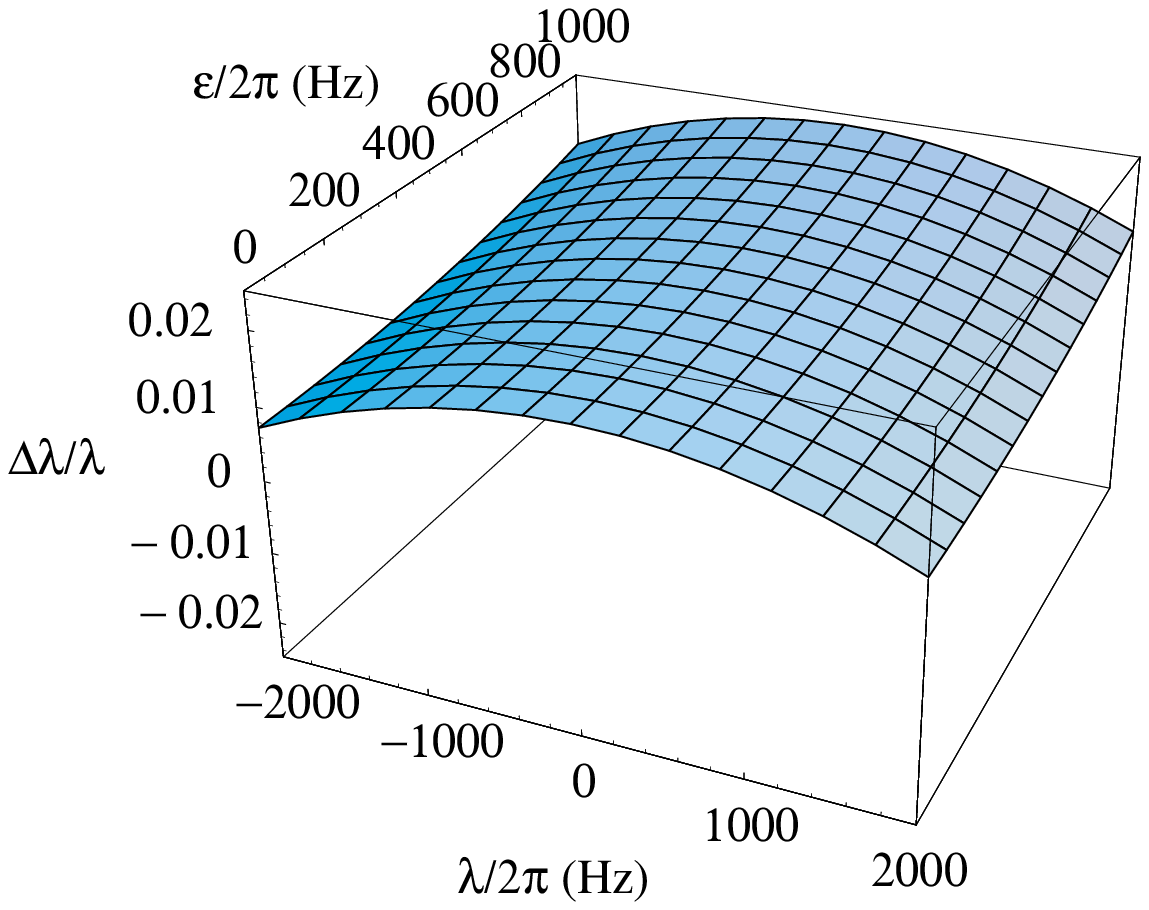,width=0.4\textwidth} & 
\epsfig{file=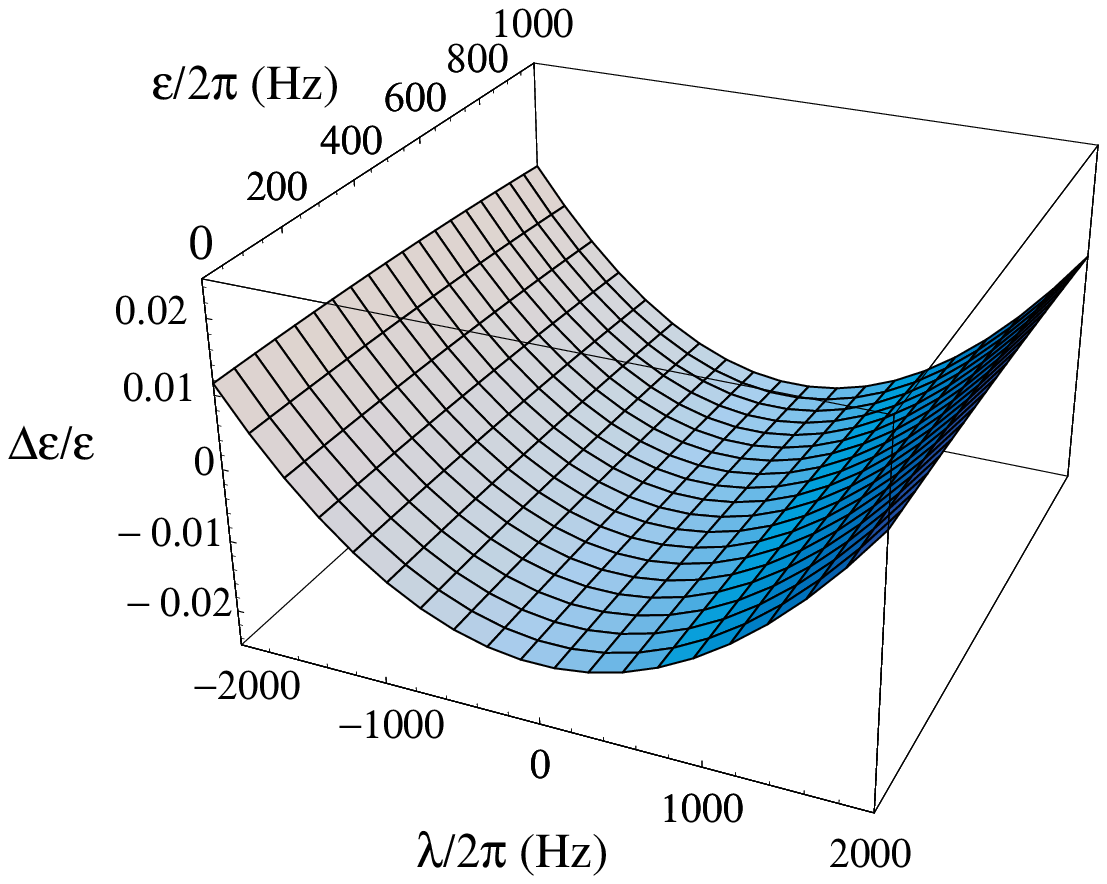,width=0.4\textwidth}
\end{tabular}
\caption{We plot the fractional error $\Delta \lambda/\lambda$ (in left panel) 
and $\Delta \epsilon/\epsilon$ (in right panel) as a function of $\lambda$ and $\epsilon$.  
The quantities $\Delta \lambda$ and $\Delta \epsilon$ are the difference between 
the value of $\lambda$ and $\epsilon$ obtained from the first-order-$T$  
 free optical frequency (\protect\ref{approxomega}) and the exact one 
(\protect\ref{exactomega}).}
\label{errorel}
\end{center}
\end{figure}
\subsection{Free optical resonant frequencies}
\label{sec5.1}

It is interesting to investigate the error in the prediction of the (free) 
optical resonant frequency introduced by using only the leading order 
terms in $T$ and $\Omega L/c$. For a  generic set of $T$, $\rho$ and $\phi$, 
it can be quite complicated to characterize that error. For example, when
$\rho>\sqrt{R}$ and $\phi\sim\pi/2$, $\widetilde{\rho}'$ 
is near $-1$ (in the complex plane) and the expansion (\ref{rhop}) around $\widetilde{\rho}'=1$ totally breaks
down. However, we are only concerned with those parameters meaningful
for a GW detector, and thus we limit our analysis to the region
where $|\widetilde{\Omega}|=\sqrt{\lambda^2+\epsilon^2}\sim\Omega_{\rm
  GW}<10^4\,\mathrm{s}^{-1}$, corresponding to 
$|\widetilde{\Omega}L/c|\stackrel{<}{_\sim} 0.1$.
In this way $|\widetilde{\rho}'-1|$ is always relatively small. 
To test the accuracy, we fix $T$, and for each $\widetilde{\Omega}=-\lambda-i\epsilon$, we
solve Eq.~(\ref{exactomega}) for $\rho$ and $\phi$. Then, 
we insert these values into Eq.~(\ref{FORF}) to get the first-order-$T$  
expression for $\widetilde{\Omega}$, which we denote by $\widetilde{\Omega}^{(1)}$. The result is: 
\bea
\label{approxomega}
\widetilde{\Omega}^{(1)}&=&
\frac{c}{L}\,\left(\frac{1+\sqrt{R}}{2}\right)^2\, \tan\left(\frac{\widetilde{\Omega}L}{c}\right)\,, \nonumber \\
&=&\widetilde{\Omega}
\left[1-\frac{T}{2}+
{\cal O}(T^2)\right]\left[1+\frac{1}{3}\left(\frac{\widetilde{\Omega}L}{c}\right)^2+{\cal O}
\left (\frac{\widetilde{\Omega}^4L^4}{c^4}\right)\right]\,.
\eea
{}From this equation we infer that since $|\widetilde{\Omega}L/c|\stackrel{<}{_\sim} 0.1$, 
and $T$ is smaller than a few percents, the error in the (free) optical resonant frequency is
not very significant (less than a few percents). In Fig.~\ref{errorel} we
plot the fractional differences (denoted by $\Delta \lambda/\lambda$ and $\Delta \epsilon/\epsilon$) 
between the real and imaginary parts of $\widetilde{\Omega}^{(1)}$ and $\widetilde{\Omega}$, as 
functions of $\epsilon$ and $\lambda$ for $T=0.033$. The fractional differences 
are always smaller than $2.5\%$.

\subsection{Input--output relation and noise spectral density}
\label{sec5.2}

Using the formalism of Sec.~\ref{sec2} and Appendix~\ref{quadraturerelations}, 
it is rather easy to derive the exact input--output relation 
in terms of $\lambda$, $\epsilon$ and $\iota_c$. 
The input--output relation ($j$-$k$) of the arm cavity composed of the effective ITM and ETM is:
\beq
\left(
\begin{array}{c}
k_1 \\k_2
\end{array}
\right)
=e^{2i\Omega L/c}
\left(
\begin{array}{cc}
1 & 0 \\
- {\cal K}_{\rm arm} & 1
\end{array}
\right)
\left(
\begin{array}{c}
j_1 \\j_2
\end{array}
\right)
+e^{i\Omega L/c}\sqrt{2{\cal K}_{\rm arm}}\frac{h}{h_{\rm SQL}^{\rm arm}}
\left(
\begin{array}{c}
 0  \\ 1
\end{array}
\right)\,,
\eeq
where
\beq
{\cal K}_{\rm arm}=\frac{8 I_{\rm arm}\omega_0}{\mu_{\rm arm} \Omega^2
  c^2}=\frac{16 I_c \omega_0}{m\Omega^2c^2}\,,\quad h_{\rm SQL}^{\rm
  arm}=\sqrt{\frac{8\hbar}{\mu_{\rm arm}\Omega^2
    L^2}}=\sqrt{\frac{8\hbar}{m\Omega^2 L^2}}\,.
\eeq
Writing Eqs.~(\ref{efftwoport}) and (\ref{efftwoport1}) in terms 
of quadratures, that is
\beq
\left(
\begin{array}{c}
{j}_1 \\{j}_2
\end{array}
\right)
=\sqrt{1-|\widetilde{\rho}'|^2}\,
\left(
\begin{array}{c}
\widetilde{a}_1 \\ \widetilde{a}_2
\end{array}
\right)
+|\widetilde{\rho}'|\,
\left(
\begin{array}{cc}
\cos \psi & -\sin \psi \\
\sin \psi  & \cos \psi
\end{array}
\right)\,
\left(
\begin{array}{c}
{k}_1 \\ {k}_2
\end{array}
\right)\,,
\eeq
and 
\beq
\left(
\begin{array}{c}
\widetilde{b}_1 \\\widetilde{b}_2
\end{array}
\right)
=\sqrt{1-|\widetilde{\rho}'|^2}\,
\left(
\begin{array}{c}
k_1 \\k_2
\end{array}
\right)
-|\widetilde{\rho}'|\,
\left(
\begin{array}{cc}
\cos \psi & \sin \psi \\
-\sin \psi  & \cos \psi
\end{array}
\right)\,
\left(
\begin{array}{c}
\widetilde{a}_1 \\ \widetilde{a}_2
\end{array}
\right)\,,
\eeq
where $\psi = {\rm arg}(\widetilde{\rho}')$, 
and using Eq.~(\ref{exactrhop}), we obtain the input--output relation 
($\widetilde{a} - \widetilde{b}$) of the three-mirror cavity, and
thus that of the equivalent SR interferometer. They can 
be represented in the same form as Eq.~(\ref{inoutsi}), 
with $\widetilde{M}^{(1)}$, $\widetilde{C}_{ij}^{(1)}$, and 
$\widetilde{D}_i^{(1)}$ replaced by:
\beq
\label{exactM}
\widetilde{M}^{\rm ex}=\frac{\Omega^2 c^2 e^{-2i\Omega L/c}}{4L^2}
\left\{
\left[1-e^{2i(\Omega+\lambda+i\epsilon)L/c}\right]
\left[1-e^{2i(\Omega-\lambda+i\epsilon)L/c}\right]
+i\frac{\iota_c L}{\Omega^2 c}
\left[e^{2i(\Omega+\lambda+i\epsilon)L/c}-e^{2i(\Omega-\lambda+i\epsilon)L/c}\right]
\right\}\,,
\eeq
and
\bea
\label{exactC11}
\widetilde{C}_{11}^{\rm ex}&=&\widetilde{C}_{22}^{\rm ex}
=\frac{\Omega^2 c^2}{4L^2}
\left\{
\left[1-2e^{-2\epsilon L/c}\cos(2\lambda
    L/c)\cos(2\Omega L/c)+e^{-4\epsilon L/c}\cos(4\lambda L/c)
  \right]+\frac{\iota_c
  L}{\Omega^2 c}e^{-4\epsilon L/c}\sin(4\lambda
  L/c) \right\}\,, \\
\label{exactC12}
\widetilde{C}_{12}^{\rm ex}
&=&\frac{\Omega^2 c^2}{4L^2}
\left\{-2e^{-2\epsilon L/c}
\sin(2\lambda L/c)
\left[\cos(2\Omega L/c)-e^{-2\epsilon L/c}\cos(2\lambda L/c)\right]
+\frac{2\iota_c L}{\Omega^2 c}e^{-4\epsilon L/c} \sin^2(2\lambda
L/c)\right\}\,, \\
\label{exactC21}
\widetilde{C}_{21}^{\rm ex}
&=&\frac{\Omega^2 c^2}{4L^2}
\left\{2
e^{-2\epsilon L/c}
\sin(2\lambda L/c) \left[\cos(2\Omega
  L/c)-e^{-2\epsilon L/c}\cos(2 \lambda L/c)\right]-\frac{2 \iota_c
  L}{\Omega^2 c}\left[1-e^{-4\epsilon L/c}\cos^2(2\lambda
  L/c)\right]\right\}\,, \\
\label{exactD1}
\widetilde{D}_{1}^{\rm ex}&=&\frac{\Omega^2 c^2}{4L^2}
\left[
-2 e^{-2\epsilon L/c}e^{i\Omega
  L/c}\sin(2\lambda L/c)\right]
 \sqrt{\frac{(1-e^{-4\epsilon L/c})\iota_c
    L}{\Omega^2 c}}\,, \\
\label{exactD2}
\widetilde{D}_{2}^{\rm ex}&=&\frac{\Omega^2 c^2}{4L^2} 
\left[2 e^{-i\Omega L/c}-2e^{-2\epsilon L/c}e^{i\Omega L/c}
  \cos(2\lambda L/c)\right]\sqrt{\frac{(1-e^{-4\epsilon L/c})\iota_c
    L}{\Omega^2 c}}\,.
\eea
\begin{figure}
\begin{center}
\begin{tabular}{cc}
\epsfig{file=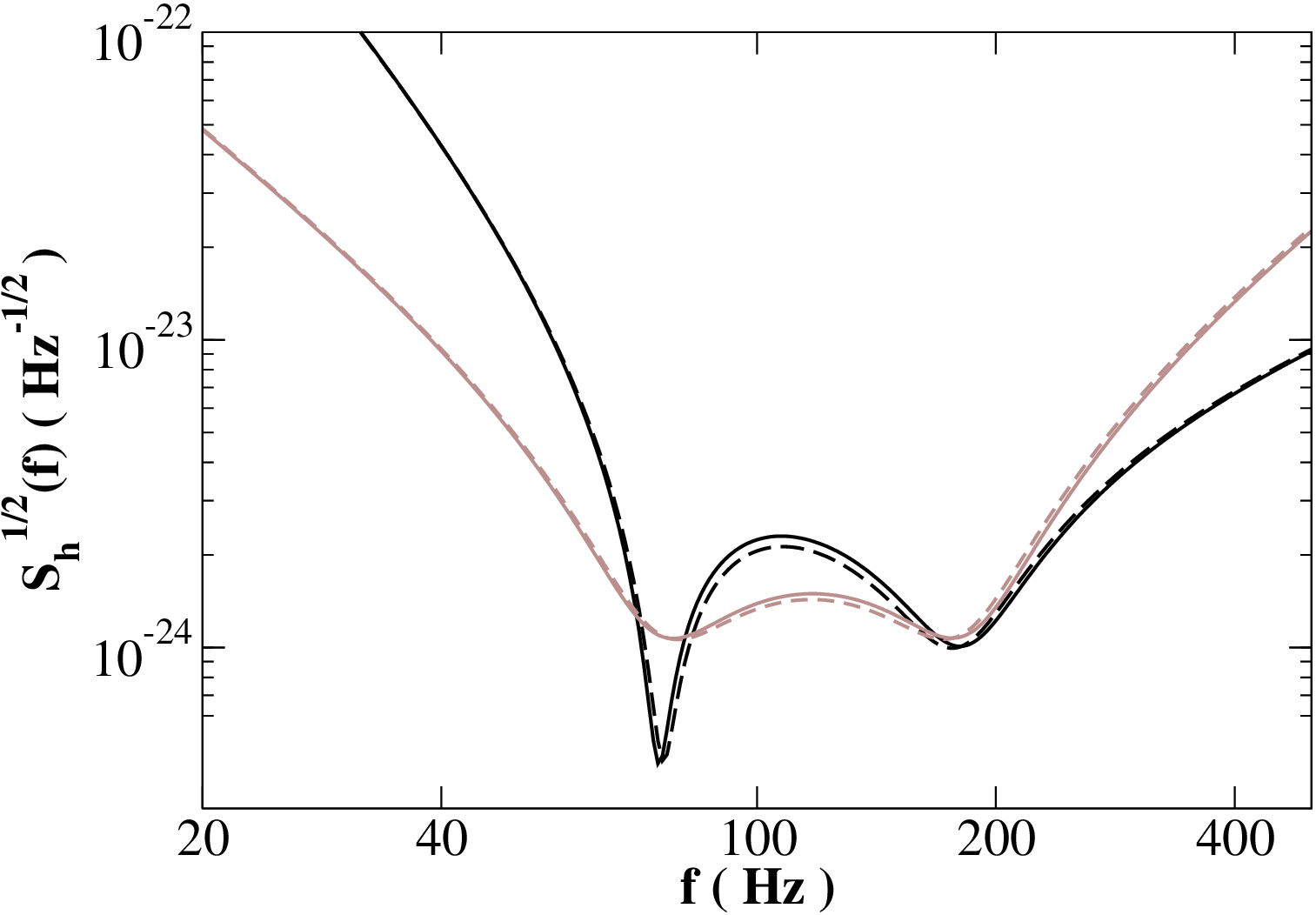,width=\sizeonefig} &
\epsfig{file=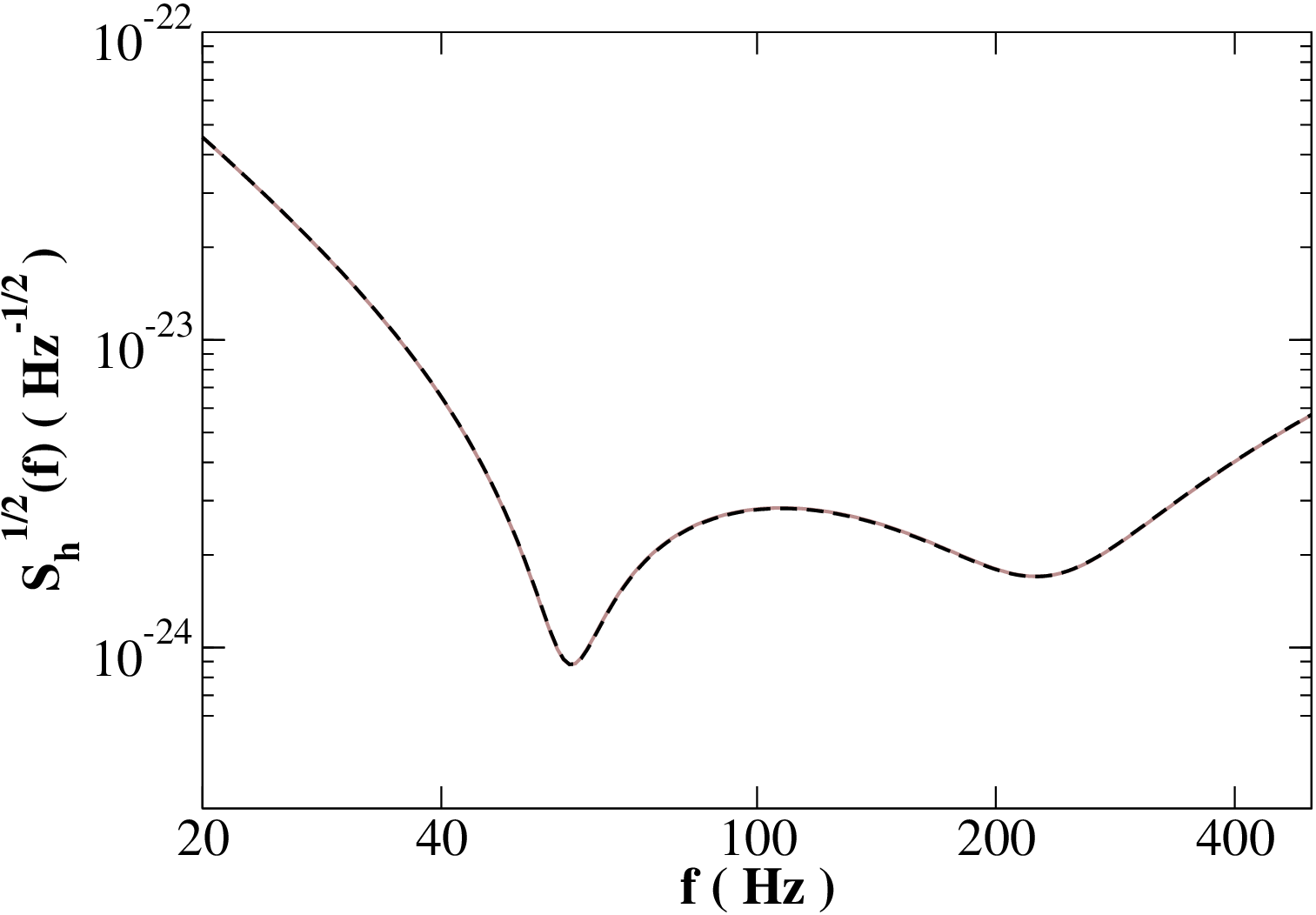,width=\sizeonefig}
\end{tabular}
\caption{\label{compareTex} 
Comparison of first-order $T$-expanded (dashed line) 
and exact (continuous line) noise spectral density 
$\sqrt{S_h}$ versus frequency $f$. 
In the left panel we use the parameters $T=0.033$, $\rho=0.9$ and
$\phi=\pi/2-0.47$, $m=30\,{\rm kg}$, and  $I_c=592\,{\rm kW}$ 
and show the curves for the two orthogonal quadratures 
$\widetilde{b}_1$ (lighter-color lines) and $\widetilde{b}_2$ 
(darker-color lines). In the right panel  we use $T=0.005$, $\rho=0.964$,
$\phi=\pi/2-0.06$, $m=40\,{\rm kg}$, $I_c=840\,{\rm kW}$, and
$\zeta=1.13\pi$.}
\end{center}
\end{figure}
In order to compare with the results obtained in
Refs.~\cite{BC1,BC2,BC3}, we have also to relate 
$\widetilde{a}$, $\widetilde{b}$ to $a$ and $b$. The exact 
transformations [to be compared with Eqs.~(\ref{t1}), (\ref{t2})] are:
\beq
\label{exredif}
\left(
\begin{array}{c}
\widetilde{a}_1 \\
\widetilde{a}_2
\end{array}
\right)
=
\frac{1}{\sqrt{1+2\rho\sqrt{R}\cos2\phi+\rho^2 R }}
\left(
\begin{array}{cc}
(1+\rho\sqrt{R})\cos\phi & -(1-\rho\sqrt{R})\sin\phi \\
(1-\rho\sqrt{R})\sin\phi & (1+\rho\sqrt{R})\cos\phi
\end{array}
\right)
\left(
\begin{array}{c}
a_1 \\
a_2
\end{array}
\right)\,,
\eeq
and
\beq
\left(
\begin{array}{c}
\widetilde{b}_1 \\
\widetilde{b}_2
\end{array}
\right)
=
\frac{1}{\sqrt{1+2\rho\sqrt{R}\cos2\phi+\rho^2 R}}
\left(
\begin{array}{cc}
(1+\rho \sqrt{R})\cos\phi & (1-\rho\sqrt{R})\sin\phi \\
-(1-\rho \sqrt{R})\sin\phi & (1+\rho\sqrt{R})\cos\phi
\end{array}
\right)
\left(
\begin{array}{c}
b_1 \\
b_2
\end{array}
\right)\,.
\eeq
As an example, we compare in the left panel of Fig.~\ref{compareTex} 
the exact and first-order $T$-expanded noise
spectral densities for the  two orthogonal quadratures $\widetilde{b}_1$ and 
$\widetilde{b}_2$, having fixed $T=0.033$, $\rho=0.9$, 
$\phi=\pi/2-0.47$, $m=30\,{\rm kg}$ and  $I_c=592\,{\rm kW}$ (which
corresponds to $I_0=I_{\rm SQL}$ at BS) as used in Refs.~\cite{BC1,BC2,BC3}. 
The $T$-expanded noise spectral density is given by Eq.~(\ref{sdlead}), 
where we used for $\lambda$, $\epsilon$ and the redefined output quadratures 
Eqs.~(\ref{lambdaepsilon}), (\ref{argtau}). 
The exact noise spectral density is obtained from 
Eq.~(\ref{nsd}) by replacing $\widetilde{M}^{(1)}$, $\widetilde{C}_{ij}^{(1)}$, and 
$\widetilde{D}_i^{(1)}$ with 
$\widetilde{M}^{\rm ex}$, $\widetilde{C}_{ij}^{\rm ex}$, and 
$\widetilde{D}_i^{\rm ex}$. From Fig.~\ref{compareTex}, we see 
that there is a discernible difference. In the right panel 
of Fig.~\ref{compareTex}, we compare 
the exact and first-order $T$-expanded noise spectral densities 
using the reference-design parameters of LIGO-II~\cite{LIGOIIref}: $T=0.005$, $\rho=0.964$,
$\phi=\pi/2-0.06$, $m=40\,{\rm kg}$, $I_c=840\,{\rm kW}$, and
$\zeta=1.13\pi$. In this case, the two curves agree nicely with each
other, presumably, because $T$ is rather small.
\begin{figure}
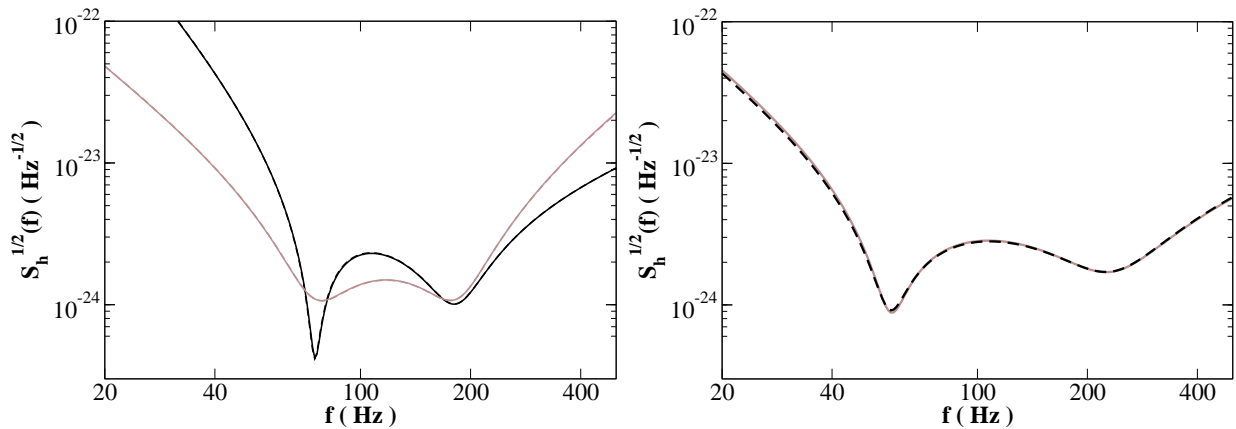

\begin{center}
\vspace{0.2cm}
\begin{tabular}{cc}
\epsfig{file=Fig9a.eps,width=\sizeonefig} &
\epsfig{file=Fig9b.eps,width=\sizeonefig}
\end{tabular}
\caption{\label{compareelex}
Comparison of first-order $\lambda$-$\epsilon$-$\iota_c^{1/3}$-expanded (dashed line) 
and exact (continuous line) noise spectral density 
$\sqrt{S_h}$ versus frequency $f$. In the left panel we use $T=0.033$, $\rho=0.9$ and
$\phi=\pi/2-0.47$, $m=30\,{\rm kg}$, and  $I_c=592\,{\rm kW}$, 
and show the curves for the two orthogonal quadratures 
$\widetilde{b}_1$ (lighter-color lines) and $\widetilde{b}_2$ (darker-color lines). 
In the right panel we use $T=0.005$, $\rho=0.964$,
$\phi=\pi/2-0.06$, $m=40\,{\rm kg}$, $I_c=840\,{\rm kW}$, and
$\zeta=1.13\pi$.}
\end{center}
\end{figure}
In the general case, if we want to trust the leading order calculation, it is not 
obvious how small $T$ can be,  since $\rho$ and $\phi$ have to change along with $T$ to
preserve the invariance of interferometer performance. For this reason, 
it is more convenient to seek an expansion that is also scaling
invariant, i.e.\ whose accuracy only depends on the scaling-invariant
properties of the interferometer. To this respect, the set of quantities $\lambda
L/c$, $\epsilon L/c$, $\iota_c^{1/3} L/c$ and $\Omega L/c$, which are
all small and on the same order, is a good choice. It is then meaningful to
expand with respect to these quantities and take the leading
order terms. We denote the noise spectral density obtained 
in this way by first-order $\lambda$-$\epsilon$-$\iota_c^{1/3}$-expanded  
noise spectral density. [This technique of identifying and expanding 
in small quantities of the same order
can be very convenient and powerful in the analysis of complicated interferometer
configurations, e.g., the speed meter interferometer~\cite{speedmeter}.]  
Not surprisingly, doing so gives us right away the scaling-invariant
input--output relation (\ref{inoutsi}). 
In the left and right panels of Fig.~\ref{compareelex} we compare 
the exact and first-order $\lambda$-$\epsilon$-$\iota_c^{1/3}$-expanded  
noise spectral densities for the  two orthogonal quadratures $\widetilde{b}_{1,2}$, 
with the same parameters used in Fig.~\ref{compareTex}, i.e.\ $T=0.033$, $\rho=0.9$ and
$\phi=\pi/2-0.47$, $m=30\,{\rm kg}$, and  $I_c=592\,{\rm kW}$ 
(left panel) and $T=0.005$, $\rho=0.964$,
$\phi=\pi/2-0.06$, $m=40\,{\rm kg}$, $I_c=840\,{\rm kW}$, and
$\zeta=1.13\pi$ (right panel). The first-order $\lambda$-$\epsilon$-$\iota_c^{1/3}$-expanded  
noise spectral density is obtained using for $\lambda$, $\epsilon$ and the redefined output 
quadratures Eqs.~(\ref{exactomega}), (\ref{exredif}). The agreement between the exact and 
first-order $\lambda$-$\epsilon$-$\iota_c^{1/3}$-expanded  noise spectral densities 
is much better than the agreement between the exact and $T$-expanded noise spectral densities, 
given in Fig.~\ref{compareTex}. 

When either $\lambda L/c$, $\epsilon L/c$, $\iota_c^{1/3} L/c$ or $\Omega L/c$ 
is not small enough, the first-order $\lambda$-$\epsilon$-$\iota_c^{1/3}$ 
expansion fails. An interesting example of astrophysical
relevance is the configuration with large $\lambda$ and small
$\epsilon$, which has narrowband sensitivities centered around a high
(optical) resonant frequency. In the left panel of Fig.~\ref{compareelexanother} we 
compare the first-order $\lambda$-$\epsilon$-$\iota_c^{1/3}$-expanded noise  
spectral density with the exact one, 
for the two quadratures $\widetilde{b}_{1,2}$ having fixed: $\lambda
= 2\pi\times 900\,{\rm Hz}$,  $\epsilon= 20\,{\rm Hz}$, $m=30\,{\rm
kg}$ and $I_c=600\,{\rm kW}$. 
Near the lower optomechanical resonant frequency, the 
first-order $\lambda$-$\epsilon$-$\iota_c^{1/3}$ expansion deviates 
from the exact one by significant amounts.  However, it is sufficient  
to expand up to the second order in $\lambda L/c$, $\epsilon L/c$, $\iota_c^{1/3} L/c$ and 
$\Omega L/c$ to get a much better agreement, 
as we infer from  the right panel of Fig.~\ref{compareelexanother}. 
[The input--output relation expanded at second order is given in 
Appendix~\ref{secondorder}.] 

\section{Conclusions}
\label{sec6}

In this paper we showed that, under the assumptions used to 
describe SR interferometers~\cite{BC1,BC2,BC3}, 
i.e.\ radiation pressure forces acting on ETMs and ITMs equal, 
and ETM and ITM motions neglected during the light round-trip time 
in arm cavities, the SR cavity can be viewed as a single effective 
(fixed) mirror located at the ITM position. We then explicitly 
map the SR optical configuration to a three-mirror cavity~\cite{JM,MR} [see 
e.g., Sec.~\ref{sec2}] or even a single detuned cavity~\cite{FK} [see 
Sec.~\ref{4.2}]. The mapping has revealed an interesting  scaling law 
present in SR interferometers. By varying the SRM reflectivity 
$\rho$, the SR detuning $\phi$ and the ITM transmissivity $T$ 
in such a way that the circulating power $I_c$ and the (free) optical 
resonant frequency (or more specifically its real and imaginary parts 
$\lambda$ and $\epsilon$) remain fixed [see Eq.~(\ref{lambdaepsilon})], 
the input--output relation and the optomechanical dynamics remain invariant.

We expressed the input--output relation (\ref{inoutsi}), noise spectral density 
(\ref{nsd}) and all quantities characterizing the optomechanical dynamics, 
such as the radiation-pressure force (\ref{rpf}) and ponderomotive rigidity (\ref{rff}), 
in terms of the scaling invariant quantities or characteristic parameters. The various formulas 
are much simpler than the ones obtained in the original description~\cite{BC1,BC2,BC3}. 
The scaling invariant formalism will be certainly useful in the process 
of optimizing the SR optical configuration of LIGO-II~\cite{BCM} and 
for investigating advanced LIGO configurations.
Moreover, the equivalance we explicitly showed between SR interferometer and 
single detuned cavity, could also make the table-top experiments of the kind recently
suggested in Ref.~\cite{BKV}  more relevant to the development of LIGO-II.
 
\begin{figure}
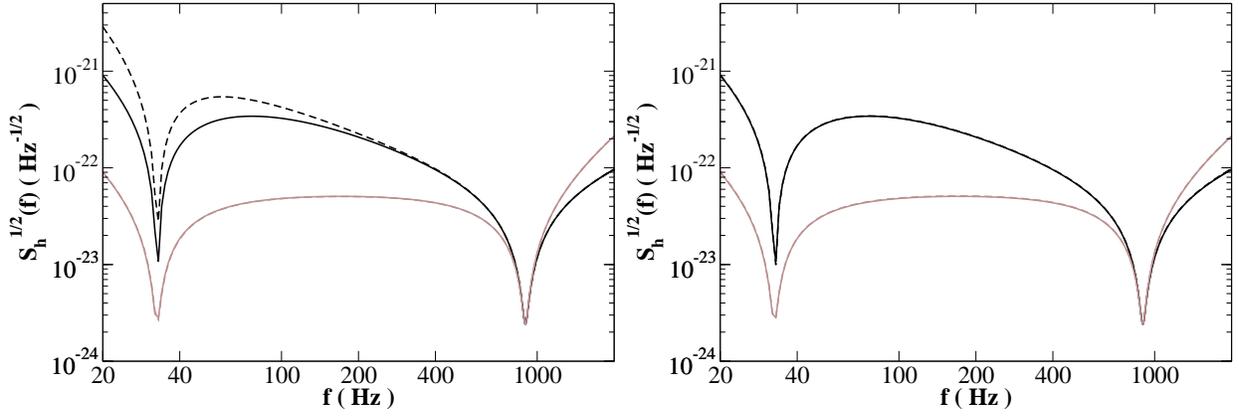

\begin{center}
\begin{tabular}{cc}
\epsfig{file=Fig10a.eps,width=\sizeonefig} &
\epsfig{file=Fig10b.eps,width=\sizeonefig}
\end{tabular}
\caption{For the two orthogonal quadratures $\widetilde{b}_1$ (lighter-color lines) 
and $\widetilde{b}_2$ (darker-color lines) we compare 
the first-order $\lambda$-$\epsilon$-$\iota_c^{1/3}$-expanded  
noise spectral density (dashed line) 
with the exact (continuous line) noise spectral density (in left panel) and 
the second-order $\lambda$-$\epsilon$-$\iota_c^{1/3}$-expanded  
noise spectral density (dashed line) with the exact (continuous line) noise spectral density 
(in right panel). For all the cases we fix $\lambda= 2\pi\times 900\,{\rm Hz}$, 
$\epsilon= 2 \pi \times 20\,{\rm Hz}$, $m=30\,{\rm kg}$ and $I_c=600\,{\rm kW}$. 
\label{compareelexanother}}
\end{center}
\end{figure}
In this paper we also evaluated the input--output relation for SR  
interferometers at all orders in the transmissivity of ITMs 
[see Sec.~\ref{sec5}]. So far, the calculations were limited 
to the leading order. We found that the differences between 
leading-order and all-order noise spectral densities 
for broadband configurations of advanced LIGO do not differ much 
[see Fig.~\ref{compareTex}]. However for narrowband configurations, which 
have an astrophysical interest, 
the differences can be quite noticeable [see left panel of 
Fig.~\ref{compareelexanother}].    
In any case, we showed that by using the (very simple) next-to-leading-order input--output 
relation, explicitly derived in Appendix~\ref{secondorder}, we can recover the 
all-order results with very high accuracy [see right panel of 
Fig.~\ref{compareelexanother}].

Finally, it will be rather interesting to investigate how the results 
change if we relax the assumption of disregarding (i) the motion of ITMs 
and ETMs during the light round-trip time in arm cavities and (ii) the radiation-pressure 
forces on ITMs due to light power present in the cavity composed of ITM and 
BS. This analysis is left for future work.

\acknowledgments 
We thank Vladimir Braginsky and Farid Khalili for stimulating discussions 
on the material presented in Sec.~\ref{4.4}, and Secs.~\ref{4.2},~\ref{4.3}, 
respectively. It is also a pleasure to thank Peter Fritschel, Nergis Mavalvala and David Shoemaker
for exchange of information on the existence of the scaling law in SR interferometers, 
and Kip Thorne for his continuous encouragement and for very useful interactions.

We acknowledge support from NSF grant PHY-0099568. 
The research for AB was also supported 
by Caltech's Richard Chace Tolman Fund. The research for YC was
also supported by the David and Barbara Groce Fund of San Diego Foundation.

\appendix{}
\section{Useful relations in the quadrature formalism}
\label{quadraturerelations}

As in Refs.~\cite{KLMTV00,BC1} we describe the interferometer's light by 
the electric field evaluated on the optic axis, i.e.\ on the 
center of light beam. Correspondingly, the electric fields that we write 
down will be functions of time only. All dependence on spatial 
position will be suppressed from our formulae.

The input field at the bright port of the beam splitter, which is assumed 
to be infinitesimally thin, is a carrier field, described by a coherent 
state with power $I_0$ and (angular) frequency $\omega_0$. 
We denote by $f_{\rm GW}=\Omega/2 \pi$ the GW frequency, which lies in 
the range $10-10^4$ Hz. The interaction of a gravitational wave with 
the optical system produces sideband frequencies $\omega_0 \pm \Omega$ 
in the electromagnetic field at the dark-port output. We describe the quantum 
optics inside the interferometer using the two-photon formalism 
developed by Caves and Schumaker \cite{CS85}. 
The quantized electromagnetic field in the Heisenberg picture 
evaluated at some fixed point on the optic axis is \cite{KLMTV00,BC1}:
\beq
E(t) = \sqrt{\frac{2 \pi \hbar\,\omega_0}{{\cal A}\,c}}\,
e^{-i\omega_0\,t}\,\int_0^{+\infty} 
(a_+(\Omega)\,e^{-i\Omega t} + a_-(\Omega)\,e^{i\Omega t})\,
\frac{d \Omega}{2 \pi}  + {\rm h.c.}\,,
\label{Efield}
\eeq
where h.c. means Hermitian conjugate and we denoted $a_+(\Omega) \equiv a_{\omega_0 + \Omega}$ 
and $a_-(\Omega) \equiv a_{\omega_0 - \Omega}$. Here ${\cal A}$ is the effective cross sectional 
area of the laser beam and $c$ is the speed of light. The annihilation and creation operators 
${a}_\pm(\Omega)$ in Eq.~(\ref{Efield}) satisfy the commutation relations:
\bea
&& [a_+, a_{+^\prime}^\dagger] = 2 \pi\,\delta(\Omega - \Omega^\prime)\,,\quad \quad 
[a_-, a_{-^\prime}^\dagger] = 2 \pi\,\delta(\Omega - \Omega^\prime)\,, \\
&& [a_+, a_{+^\prime}] = 0 =  [a_-, a_{-^\prime}]\,, \quad \quad 
[a_+^\dagger, a_{+^\prime}^\dagger] = 0 =  [a_-^\dagger, a_{-^\prime}^\dagger]\,,
\quad \quad  [a_+, a_{-^\prime}] = 0 = [a_+, a_{-^\prime}^\dagger]\,.
\eea
Following the Caves-Schumaker two-photon formalism~\cite{CS85}, we introduce 
the amplitudes of the two-photon modes as  
\beq
a_1 = \frac{a_+ + a_-^\dagger}{\sqrt{2}}\,, \quad \quad 
a_2 = \frac{a_+ - a_-^\dagger}{\sqrt{2}i}\,;
\label{quadr}
\eeq
$a_1$ and $a_2$ are called quadrature fields and they satisfy the 
commutation relations:
\bea
&& [a_1, a_{2^\prime}^{\dagger}] = - [a_2, a_{1^\prime}^{\dagger}]=
2\pi i \delta(\Omega-\Omega^\prime)\,, \nonumber \\
&& [a_1, a_{1^\prime}^{\dagger}] = 0= [a_1, a_{1^\prime}]\,,
\quad \quad 
[a_2, a_{2^\prime}^{\dagger}] = 0= [a_2, a_{2^\prime}]\,.
\eea
The electric field (\ref{Efield}) in terms of the quadratures reads:
\beq 
E(a_i;t) = \cos (\omega_0\,t)\,E_1(a_1;t) + \sin (\omega_0\,t)\,E_2(a_2;t)\,,
\eeq
where: 
\beq
E_j(a_j;t) = \sqrt{\frac{4 \pi \hbar\,\omega_0}{{\cal A}\,c}}\,
\int_0^{+\infty} (a_j\,e^{-i\Omega t} + a_j^{\dagger}\,e^{i\Omega t})\,
\frac{d \Omega}{2 \pi} \quad \quad j = 1,2\,.
\eeq
Any linear relation among the fields $a_\pm(\Omega)$ of the kind: 
\beq
\label{bpm-apm}
b_\pm{(\Omega)}=f_\pm(\Omega)\,a_\pm{(\Omega)}\,, \quad \quad 
f_+(\Omega) \equiv f(\omega_0+\Omega)\,, \quad \quad f_-(\Omega) \equiv f(\omega_0-\Omega)\,,
\eeq
can be transformed into the following relation among the quadrature fields:
\beq
\left(
\begin{array}{c}
\label{generalquadrature}
b_1 \\ b_2
\end{array}
\right)
=\frac{1}{2}
\left(
\begin{array}{cc}
(f_++f_-^*) & i\,(f_+-f_-^*) \\
-i\,(f_++f_-^*) & (f_++f_-^*) 
\end{array}
\right)
\left(
\begin{array}{c}a_1 \\ a_2
\end{array}
\right)\,.
\eeq
In general, the above equation can be very complicated. In this paper we restrict
ourselves to two  special cases.  
The first case is when $|f_+|=|f_-|$ and we write
\beq
\label{FPsi}
f_{\pm}(\Omega) =F(\Omega)\, e^{i \Psi_{\pm}(\Omega)}\,\quad\forall\, \Omega\, >0\,,
\eeq
and Eq.~(\ref{generalquadrature}) becomes:
\beq
\label{quadraturespecial}
\left(
\begin{array}{c}
b_1 \\ b_2
\end{array}
\right)
=
F(\Omega)\, e^{i(\Psi_+-\Psi_-)/2}
\left(
\begin{array}{rr}
\cos\frac{\Psi_++\Psi_-}{2} & -\sin\frac{\Psi_++\Psi_-}{2} \\
\sin\frac{\Psi_++\Psi_-}{2} & \cos\frac{\Psi_++\Psi_-}{2}
\end{array}
\right)
\left(
\begin{array}{c}a_1 \\ a_2
\end{array}
\right)\,.
\eeq
It is easily checked that the input--output relation for the following 
processes: (i) free propagation in space, (ii) 
reflection and transmission from a thin mirror, (iii)
reflection and transmission from one (or more) Fabry-Perot cavity 
for which $\omega_0$ is either resonant or antiresonant, and 
(iv) reflection and transmission from one (or more) FP cavity 
whose bandwidth is much larger than the range of values 
of $\Omega$ we are interested in [in this case 
$f(\Omega)$ can be considered as a constant (complex) number]
are all special cases (or linear combinations) of the relation (\ref{quadraturespecial}).

The second case of interest for us is when there is \emph{one} resonance at
$\omega_0+\Omega_{\rm r}$, with $\Omega_{\rm r}$ complex. 
In this case $f(\Omega)$ is of the form:
\beq
\label{ares}
f(\omega)= \frac{g(\omega)}{\omega-\omega_0-\Omega_{\rm r}}\,,
\eeq
where $g(\omega)$ does not have poles. For $\Omega>0$, we have
\beq
f_+=\frac{g(\omega_0+\Omega)}{\Omega-\Omega_{\rm r}}\,, 
\quad \quad f_-^*=-\frac{g^*(\omega_0-\Omega)}{\Omega+\Omega^*_{\rm r}}\,,
\eeq
and thus 
\bea
f_++f_-^*&=&\frac{(\Omega+\Omega^*_{\rm r})\,        
g(\omega_0+\Omega)-(\Omega-\Omega_{\rm r})\,g^*(\omega_0-\Omega)}
{(\Omega-\Omega_{\rm r})\,(\Omega+\Omega^*_{\rm r})}\,, \\
f_+-f_-^*&=&\frac{(\Omega+\Omega^*_{\rm r})\, 
g(\omega_0+\Omega)+(\Omega-\Omega_{\rm r})\,g^*(\omega_0-\Omega)}
{(\Omega-\Omega_{\rm r})\,(\Omega+\Omega^*_{\rm r})}\,.
\eea
Since the quadrature field at $\Omega$ mixes the frequencies $\omega_0+\Omega$ and $\omega_0-\Omega$, 
the single resonant frequency $\Omega_{\rm r}$ appears in the above equation 
as a pair of resonant frequencies $\{\Omega_{\rm r}, -\Omega^*_{\rm r}\}$.

\section{The Stokes relations}
\label{stokes}

\begin{figure}
\begin{center}
\epsfig{file=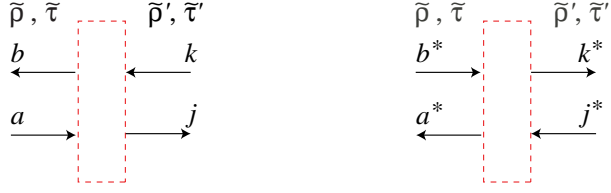,width=\sizeonefig}
\caption{\label{stokesfig} A two-port linear optical 
system can always be expressed in terms of four effective 
transmissivities and reflectivities, $\widetilde{\rho}'$, 
$\widetilde{\tau}'$ (for fields entering from the right side),  
and $\widetilde{\rho}$, $\widetilde{\tau}$ (for fields entering from 
the left side). By taking the complex conjugates
of the field amplitudes and inverting 
their propagation directions, a new set of fields related by the same set 
of transmissivities and reflectivities is obtained.}
\end{center}
\end{figure}

The transmission and reflection coefficients of 
a system of mirrors, or more generally of a two-port linear optical 
system, can always be expressed in terms of four effective 
transmissivities and reflectivities:
$\widetilde{\rho}$, $\widetilde{\tau}$, $\widetilde{\rho}'$ and
$\widetilde{\tau}'$ [see Fig.~\ref{stokesfig}]. These quantities are generally frequency 
dependent (complex) numbers. For the fields shown in Fig.~\ref{stokesfig}, we have:
\bea
j_{\omega}&=&\widetilde{\rho}' k_{\omega}+\widetilde{\tau} a_{\omega}\,, \\
b_{\omega}&=&\widetilde{\tau}' k_{\omega}+\widetilde{\rho} a_{\omega}\,.
\eea
Imposing that the two-port linear optical system satisfies 
the conservation of energy, we have:
\beq
\label{stoke0}
|\widetilde{\rho}|^2+|\widetilde{\tau}|^2=1\,,\quad 
|\widetilde{\rho}'|^2+|\widetilde{\tau}'|^2=1\,.
\eeq
If we take the complex conjugates of all the complex amplitudes and revert their 
propagation directions, the resulting configuration is also a solution of the optical system, 
in the sense that the new fields are also related by the same sets of effective 
transmissivities and reflectivities. Thus, the system is invariant under time reversal.  
By applying explicitly this symmetry, it is straightforward to derive:
\bea
\label{stoke1}
&& \widetilde{\rho}\,\widetilde{\rho}^*+\widetilde{\tau}'\,\widetilde{\tau}^*=1\,, \quad \quad 
\widetilde{\rho}^*\,\widetilde{\tau}+\widetilde{\tau}^*\,\widetilde{\rho}'=0\,,\\
\label{stoke2}
&& \widetilde{\rho}'\,\widetilde{\rho}^{'*}+\widetilde{\tau}\,\widetilde{\tau}^{'*}= 1\,, 
\quad \quad \widetilde{\rho}^{'*}\,\widetilde{\tau}'+\widetilde{\tau}^{'*}\,\widetilde{\rho} =0\,.\\
\eea
Equations (\ref{stoke0})--(\ref{stoke2}) are the well-known Stokes
relations~\cite{BK}.  If we rewrite the transmissivity and reflectivity coefficients as
\bea
\widetilde{\rho}=|\widetilde{\rho}|\, 
e^{i\mu}\,, && \widetilde{\tau}=|\widetilde{\tau}|\, e^{i\nu}\,, \\
\widetilde{\rho}'=|\widetilde{\rho}'|\,e^{i\mu'}\,, 
&& \widetilde{\tau}'=|\widetilde{\tau}'|\,e^{i\nu'}\,,
\eea
and insert them into the Stokes relations (\ref{stoke1})--(\ref{stoke2}), we obtain 
\bea
\label{stokeap1}
&& |\widetilde{\rho}|=|\widetilde{\rho}'|\,, \quad \quad |\widetilde{\tau}|=|\widetilde{\tau}'|\,,
\quad \quad |\widetilde{\rho}|^2+|\widetilde{\tau}|^2=1\,;\\
\label{stokeap5}
&& e^{i \nu}=e^{i \nu'}\,,\quad \quad e^{i(\mu+\mu')}=-e^{2 i \nu}\,.
\eea

\section{Input--output relations at second order in transmissivity of internal test masses}
\label{secondorder}

The input--output relation expanded up to second order in 
$\lambda L/c$, $\epsilon L/c$, $\iota_c^{1/3} L/c$ and $\Omega L/c$
can be obtained in a straightforward way by expanding 
Eqs.~(\ref{exactM})--(\ref{exactD2}). The new 
coefficients $\widetilde{M}^{(2)}$,
$\widetilde{C}^{(2)}_{ij}$ and $\widetilde{D}^{(2)}_i$ are very simple. In fact, they can be
represented in terms of the first-order ones, $\widetilde{M}^{(1)}$,
$\widetilde{C}^{(1)}_{ij}$ and $\widetilde{D}^{(1)}_i$ 
given by Eqs.~(\ref{denm})--(\ref{coeffd}), through the following formulas
(truncated at the next-to-leading order): 
\beq
\widetilde{M}^{(2)}=(1-2\epsilon L/c)\widetilde{M}^{(1)}\,,
\eeq
\beq
\left(
\begin{array}{cc}
\widetilde{C}_{11}^{(2)} & \widetilde{C}_{12}^{(2)} \\
\widetilde{C}_{21}^{(2)} & \widetilde{C}_{22}^{(2)}
\end{array}
\right)
=(1-2\epsilon L/c) 
\left(
\begin{array}{cc}
1 & \lambda L/c \\
-\lambda L/c & 1
\end{array}
\right)
\left(
\begin{array}{cc}
\widetilde{C}_{11}^{(1)} & \widetilde{C}_{12}^{(1)} \\
\widetilde{C}_{21}^{(1)} & \widetilde{C}_{22}^{(1)}
\end{array}
\right)
\left(
\begin{array}{cc}
1 & \lambda L/c \\
-\lambda L/c & 1
\end{array}
\right)\,,
\eeq
and
\beq
\left(
\begin{array}{c}
\widetilde{D}_1^{(2)} \\
\widetilde{D}_2^{(2)}
\end{array}
\right)
=(1-2\epsilon L/c)
\left(
\begin{array}{cc}
1 & \lambda L/c \\
-\lambda L/c & 1
\end{array}
\right)
\left(
\begin{array}{c}
\widetilde{D}_1^{(1)} \\
\widetilde{D}_2^{(1)}
\end{array}
\right)\,.
\eeq
It is quite remarkable that, at second order, the 
optomechanical resonances, determined by $\widetilde{M}^{(2)}=0$, remain
unchanged with respect to the first order result obtained imposing 
$\widetilde{M}^{(1)}=0$. Apart from a (frequency-independent) rotation 
of the quadrature phases, 
the input--output relation at next-to-leading order are very similar 
to the leading-order one.

\end{document}